\documentclass[english,journal=jctcce,manuscript=article,articletitle=true,layout=twocolumn]{achemso}
\setkeys{acs}{articletitle=true}
\setkeys{acs}{doi = false} 
\setkeys{acs}{maxauthors=10}
\setkeys{acs}{etalmode=truncate}

\usepackage[english]{babel}
\usepackage[T1]{fontenc}

\usepackage{amsmath}
\usepackage{amssymb}
\usepackage{graphicx}
\usepackage{float}
\usepackage[colorlinks=true, allcolors=blue]{hyperref}
\usepackage{cleveref}
\usepackage[version=4]{mhchem}
\usepackage[table]{xcolor}
\usepackage{rotating}
\usepackage{caption}
\usepackage{subcaption}
\usepackage{makecell}

\newcommand*\citeit[1]{\citeauthor{#1}\cite{#1}}

\newcommand*\HelFEM{\textsc{HelFEM}}
\newcommand*\Libxc{\textsc{Libxc}}

\newcommand*\citeref[1]{ref.~\citenum{#1}}
\newcommand*\citerefs[1]{refs.~\citenum{#1}}

\title{Systematic Study of Confinement Induced Effects on Atomic Electronic Structure}

\author{Hugo {\AA}str{\"o}m}
\affiliation{University of Helsinki, Department of Chemistry, 
Faculty of Science, P.O. Box
  55 (A.I. Virtanens plats 1), FI-00014 University of Helsinki, Finland}
\author{Susi Lehtola}
\email{susi.lehtola@alumni.helsinki.fi}
\affiliation{University of Helsinki, Department of Chemistry, 
Faculty of Science, P.O. Box
  55 (A.I. Virtanens plats 1), FI-00014 University of Helsinki, Finland}

\begin{document}

\begin{abstract}
  We point out that although a litany of studies have been published on atoms in hard-wall confinement, they have either not been systematic, having only looked at select atoms and/or select electron configurations, or they have not used robust numerical methods.

  To remedy the situation, we perform in this work a methodical study of atoms in hard-wall confinement with the \textsc{HelFEM} program, which employs the finite element method (FEM) that trivially implements the hard-wall potential, guarantees variational results, and allows easily finding the numerically exact solution.

  Our fully numerical calculations are based on non-relativistic density functional theory (DFT) and spherically averaged densities.
   We consider three levels of density functional approximations: the local density approximation employing the Perdew--Wang (PW92) functional, the generalized-gradient approximation (GGA) employing the Perdew--Burke--Ernzerhof (PBE) functional, and the meta-GGA approximation employing the r$^2$SCAN functional.
   Importantly, the completely dissimilar density functional approximations are in excellent agreement, suggesting that the observed results are not artefacts of the employed level of theory.

  We systematically examine low lying configurations of the H--Xe atoms and their monocations, and investigate how the configurations---especially the ground state configuration---behave as a function of the position of the hard-wall boundary.
  We perform calculations with both spin-polarized as well as spin-restricted densities, and demonstrate that spin-polarization effects are significant in open shell configurations, even though some previous studies have only considered the spin-restricted model.

  We demonstrate the importance of considering ground state changes for confined atoms by computing the ionization radii for the H--Xe atoms and observe significant differences to earlier studies.
  Confirming previous observations, we identify electron shifts on the outermost shells for a majority of the elements: valence $s$ electrons are highly unfavored under strong confinement, and the high-lying $3d$ and $4f$ orbitals become occupied in atoms of periods 2--3 and 3--4, respectively.

  We also comment on deficiencies of a commonly used density based estimate for the van der Waals (vdW) radius of atoms, and propose a better behaved variant in terms of the number of electrons outside the vdW radius that we expect will prove useful in future studies.
\end{abstract}

\section{Introduction} \label{sec:intro}

The pioneering study of \citeit{Michels1937_P_981} was concerned on the question of how atomic polarizabilities---which have an important role in chemical bonding---evolve as a function of pressure.
Published in \citeyear{Michels1937_P_981}, the study examined the hydrogen atom enclosed in an impenetrable sphere of a variable radius to simulate pressure effects.
Such frugal models offer a great tool for understanding basic physical effects precisely due to their simplicity: \citeit{Michels1937_P_981} found that hydrogen becomes less polarizable in increasing pressure, that is, when the radius of the sphere is decreased .

An interest in analogous studies of atoms in hard-wall confinement (i.e. atoms in a spherical box) continues to this day.
A multitude of studies have been published on the very first atoms of the periodic table; see \citeref{Aquino2009_AQC_123} for a review dedicated to the H and He atoms confined by a finite or infinite spherical barrier with 200 references.
Yet, few studies have considered many-electron atoms, or reported systematic calculations on the periodic table.

In hard-wall confinement, the wave function must vanish beyond the hard-wall boundary, and this can only be accomplished within a basis set with finite support.
Many studies on confined atoms employing basis sets of an analytic form---such as Gaussian or Slater type orbital basis sets---have been published in the literature.
However, these basis sets have finite tails in the forbidden region.
Truncated Gaussian or Slater type orbital basis sets where this tail has been cut off do not have this issue, but they too suffer from basis set truncation errors (BSTEs), which are \emph{a priori} unknown.

Our recent studies on atoms and molecules in strong magnetic fields found that changes in the electron configuration often resulted in significant increases in the BSTE of standard Gaussian-type orbital basis sets,\cite{Lehtola2020_MP_1597989, Aastroem2023_JPCA_10872} as these basis sets have not been optimized to describe electronic structure in this context.
As we will discuss below, confinement is likewise expected to result in electron shifts to orbitals of higher angular momentum.
Since this effect has again not been considered in the parametrization of standard Gaussian-type orbital basis sets, they cannot be expected to yield reliable estimates under confinement, since the orbitals of confined atoms can look extremely dissimilar to the orbitals of the unconfined atom.
The qualitatively correct description of confinement induced effects then requires a careful study of basis set convergence, ideally supplemented by comparisons to fully numerical reference values.
We point out that few studies have attempted to reach the complete basis set limit, and we will report numerically exact complete basis set limit energies later on in this work.

Before laying out the scope of the present study, we will briefly review the literature we found that fulfills our key criteria at least in part: studies mentioned herein have examined atoms with more than 4 electrons, and employed a robust numerical method.

\citeit{Boeyens1994_JCSFT_3377} performed fully numerical Hartree--Fock calculations to determine the ionization radii of compressed atoms.
However, \citeit{Boeyens1994_JCSFT_3377} did not employ a proper hard-wall potential.
To approximate a hard-wall boundary at $r=r_\mathrm{c}$, \citeit{Boeyens1994_JCSFT_3377} multiplied the wave function by the smooth step function $\Theta(r) = \exp \left[ -\left( r/r_c \right)^{20} \right]$ at every self-consistent field (SCF) iteration, before the normalization of the updated wave function was carried out.
\citeit{Boeyens1994_JCSFT_3377} employed standard electron configurations for the elements, and thus did not consider changes in the occupations as a function of the confinement.

\citeit{Chattaraj2003_CPL_805} employed a similar approach to that of \citeit{Boeyens1994_JCSFT_3377} in a density functional theory (DFT) study on chemical reactivity indices of the He--Ne and Br atoms, finding that the atoms become harder and less polarizable in increasing confinement, thus coming to the same conclusion as \citeit{Michels1937_P_981} in the case of hydrogen 66 years earlier.
\citeit{Chattaraj2003_JPCA_4877} also carried out a similar study for the reactivity indices for the He--Ne atoms and the \ce{C+}, \ce{C^{2+}}, \ce{C^{3+}}, and \ce{C^{4+}} ions, with the same conclusion.
Yet, it appears that \citeauthor{Chattaraj2003_CPL_805}  did not consider the possibility of the electron configuration changing as a function of confinement in \citerefs{Chattaraj2003_CPL_805} and \citenum{Chattaraj2003_JPCA_4877}.

\citeit{Sarkar2009_JPCA_10759} studied the effects of confinement on the chemical reactivity of the nitrogen atom.
\citeit{Sarkar2009_JPCA_10759} compared the method of \citeit{Boeyens1994_JCSFT_3377} to the true hard-wall potential, and found the arising chemical reactivities to yield the same qualitative trends.
Moreover, \citeit{Sarkar2009_JPCA_10759} found that different methods to calculate the reactivity yield dissimilar results, and that different electronic occupations also lead to different chemical reactivities.

\citeit{Connerade2000_JPBAMOP_251} studied the filling of shells in the $3d$ and $4d$ elements as well as their cations in average-of-configurations Hartree--Fock calculations in a spherical cavity surrounded by a $10 E_h$ potential wall.
They found that  orbital filling becomes more regular for successive rows with increasing pressure, that is, when the finite barrier is placed closer and closer to the nucleus.
In specific, the $s$--$d$ competition disappears for these elements, as the $(n-1)d$ orbitals become energetically more favourable than the $ns$ orbitals under confinement.
Connerade and coworkers have also reported studies on some individual atoms under confinement.
For instance, \citeit{Connerade1998_JPBAMOP_3557} examined the Cr atom, while \citeauthor{Connerade2000_JPBAMOP_869} performed relativistic calculations on the La and Cs atoms in \citerefs{Connerade2000_JPBAMOP_869} and \citenum{Connerade2000_JPBAMOP_3467}, respectively.

A number of studies of atoms in hard-wall confinement have been carried out at the DFT level by \citeauthor{Garza1998_PRE_3949}.
\citeit{Garza1998_PRE_3949} reported a finite differences method (FDM) implementation for solving the atomic DFT equations in the presence of a hard-wall boundary, and studied the Ne and Na (and He) atoms in exchange-only calculations.
In a follow-up study, \citeit{Garza2000_JMST_183} examined the electrostatic potential of the unconfined and hard-wall confined Li, Na, K, and Rb atoms with DFT in order to study changes in the atoms' shell structure, and found that shell structure is gradually lost in increasing confinement.
Similarly, \citeit{Sen2000_CPL_29} determined atomic ionization radii for the He--Ca atoms with hard-wall confined DFT calculations.
As none of the studies by \citeit{Garza1998_PRE_3949}, \citeit{Garza2000_JMST_183}, or \citeit{Sen2000_CPL_29} discuss occupations, we can only assume that the ground state electron configuration of the unconfined atom was used in their calculations.
In contrast, the study on the chemical reactivity indices of the compressed Li, Na, and K; Be, Mg, and Ca; N, P, and As; and Ne, Ar, and Kr atoms by \citeit{Garza2005_JCS_379}, which was likewise carried out with hard-wall confined DFT, explicitly mentions finding the lowest electron configuration for each confinement radius.

Continuing with the Garza group, \citeit{Guerra2009_AQC_1} modeled pressure effects on the electronic properties of Ca, Sr, and Ba with hard-wall confined DFT, and found that their ground state changes from the $ns^2$ configuration to the $ns^1 (n-1)d^1$ and $ns^0 (n-1)d^2$ configurations in increasing pressure.
Confinement effects on the spin potential of first-row transition metal cations were investigated at the hard-wall confined DFT level of theory by \citeit{LozanoEspinosa2017_PM_284}.

\citeit{Pasteka2020_MP_1730989} studied the C and K (and He) atoms in hard-wall confinement with the finite element method (FEM) at the complete active space self-consistent field\cite{Roos1980_IJQC_175, Roos1980_CP_157} (CASSCF) level of theory, which represents the most sophisticated level of theory in the studies we found.
Also \citeit{Pasteka2020_MP_1730989} found changes in atomic ground state configurations under confinement: the ground state configuration of carbon changes from $1s^2 2s^2 2p^2$ in the free atom to $1s^2 2p^4$ in strong confinement.

As this literature overview demonstrates, although several fully numerical studies have been published on atoms in confinement, they have mostly focused on small groups of similar atoms.
The study of \citeit{Boeyens1994_JCSFT_3377} was the most complete, considering non-relativistic calculations on the first 105 atoms of the periodic table, but the study relied on an \emph{ad hoc} implementation of confinement, and---like in many other studies mentioned above---changes in the ground state as a function of confinement were not considered.
Studies that considered changes in the ground state are likewise limited: \citeit{Connerade2000_JPBAMOP_251} only considered $3d$ and $4d$ elements, \citeit{Garza2005_JCS_379} studied only atoms of groups 1, 2, 15, and 18, and \citeit{Guerra2009_AQC_1} only studied heavy atoms of group 2.

The only study that appears to have systematically investigated effects of confinement on the electronic structure of atoms so far is the study of \citeit{Rahm2019_JACS_10253}, which employed the extreme pressure polarizable continuum model\cite{Cammi2015_JCC_2246} (XP-PCM) to study the first 93 atoms under pressure, employing the ANO-RCC Gaussian basis set\cite{Widmark1990_TCA_291, *Roos2004_TCA_345, *Roos2004_JPCA_2851, *Roos2005_JPCA_9, *Roos2005_CPL_295, *Roos2008_JPCA_11431} with scalar relativistic corrections and the 25\% hybrid\cite{Ernzerhof1999_JCP_5029, Adamo1999_JCP_6158} of the Perdew--Burke--Ernzerhof (PBE) functional,\cite{Perdew1996_PRL_3865, Perdew1997_PRL_1396} which is commonly known as PBE0.
\citeit{Rahm2019_JACS_10253} identified electron shifts, $s \to p$, $s \to d$, $s \to f$, and $d \to f$, as essential chemical and physical consequences of compression.
In follow-up work, \citeit{Rahm2020_C_2441} studied non-bonded radii of compressed atoms with analogous calculations.
\citeit{Rahm2019_JACS_10253} and \citeit{Rahm2020_C_2441} did not report carrying out a basis set convergence study of their findings.
Yet, as already discussed above, Gaussian basis sets cannot be expected to deliver reliable results in strong confinement.\cite{Pasteka2020_MP_1730989}

As is clear from the above discussions, there is a hole in the literature regarding systematic fully numerical studies covering (a significant fraction of) the periodic table.
Changes in the electronic structure of atoms are an expected effect of confinement, and they thus need to be taken into account by considering changes in the electronic occupations as a function of confinement.
Fully numerical techniques are critially important as changes in the electronic occupations cannot be expected to be adequately reproduced by standard Gaussian-type orbital basis sets, for example.

In this work, we will address this major shortcoming of the existing literature with a systematic study on atoms in hard-wall confinement within a fully numerical approach, while also thoroughly considering the effect of the confinement on the ground state electron configuration.

Our results confirm previous partial findings of the fully numerical studies of \citeit{Connerade2000_JPBAMOP_251},  \citeit{Guerra2009_AQC_1}, and \citeit{Pasteka2020_MP_1730989} as well as the thorough Gaussian basis study of \citeit{Rahm2019_JACS_10253}: confinement induced electron shifts can be observed for a majority of the elements.
It is important to note here that all of these studies have been carried out with different methodologies.
Furthermore, as we shall see, the dissimilar functionals we employ in this work are all in good agreement.
This suggests that the findings of this work indeed correspond to physical effects of confinement despite the arguably simplistic level of theory employed herein.

The layout of this work is the following.
The theory behind the present work is briefly discussed in \cref{sec:theory}.
The form of the Hamiltonian including the hard-wall potential and the employed DFT approach is discussed in \cref{sec:dft}.
The employed theories for calculating ionization energies and ionization radii are discussed in \cref{sec:ion_e,sec:ion_r_theory}, respectively.
As our study includes data on a variety of electron configurations, we discuss ways to estimate the corresponding atomic radii in \cref{sec:atomic-radii}, where we also introduce a new way to estimate atoms' van der Waals (vdW) radii.
Next, the computational details including the employed numerical approach are presented in \cref{sec:computational-details}.
The results are presented in \cref{sec:results}.
We first demonstrate that our method works by computing ionization energies and ionization radii in \cref{sec:ion_e_2,sec:ion_radii}, respectively.
Then, we analyze the numerical and physical behavior of the employed vdW radius estimates in \cref{sec:radii-results} and find our vdW estimate to address many issues in a commonly used estimate.
The highlights from our analysis of the behavior of the low lying configurations of the H--Xe atoms in hard-wall confinement are presented in \cref{sec:changes}.
The article concludes in a summary and discussion in \cref{sec:summary}.
Atomic units are employed throughout the work.
The full set of raw data as well as a detailed analysis
of the results is included in the Supporting
Information (SI).

\section{Theory} \label{sec:theory}

\subsection{DFT approach} \label{sec:dft}
The Hamiltonian for the confined atoms reads
\begin{equation}\label{eqn:hamiltonian}
\hat{H} = \hat{H}_0 + V_c(r),
\end{equation}
where
\begin{equation}
  \hat{H}_0 = -\frac 1 2 \sum_i \nabla_i^2 - \sum_{i} \frac Z {r_i} + \sum_{i<j} \frac 1 {|{\bf r}_i-{\bf r}_j|}
\end{equation}
is the standard electronic Hamiltonian for an atom with atomic number $Z$, and $V_c(r)$ is the potential for a hard-wall boundary at $r_c$
\begin{equation} \label{eqn:hard-wall}
V_c(r) = \begin{cases}
0,\quad r<r_c \\
\infty,\quad r\geq r_c.
\end{cases}
\end{equation}

As already mentioned in \cref{sec:intro}, we carry out calculations on many-electron atoms within the context of DFT.\cite{Hohenberg1964_PR_864, Kohn1965_PR_1133}
We expand the one-particle states also known as atomic orbitals (AOs) in a numerical basis set as
\begin{equation} \label{eqn:ao}
\chi^\sigma_{nlm}(\mathbf{r})=R^\sigma_{nl}(r)Y_{lm}(\hat{\mathbf{r}}),
\end{equation}
where $Y_{lm}(\hat{\mathbf{r}})$ are spherical harmonics; we
will discuss the form of the radial expansion $R^\sigma_{nl}(r)$ below in
\cref{sec:computational-details}.

Following standard practice in numerical DFT calculations on atoms, we assume that the electron density is spherically symmetric, $n_\sigma({\bf r})=n_\sigma(r)$ for each spin $\sigma$.
This simplification leads to an efficient algorithm, since the problem of solving the Kohn--Sham equations,\cite{Kohn1965_PR_1133} i.e., the optimization of the $s$, $p$, $d$, and $f$ orbitals becomes a set of coupled one-dimensional problems.\cite{Lehtola2020_PRA_12516, Lehtola2023_JCTC_2502}

Furthermore, we will assume integer occupations on the $s$, $p$, $d$ and $f$ shells in the aim to extract chemical understanding from the calculations.
These integer occupations are then averaged over all magnetic sublevels to produce spherically symmetric total densities; this leads to fractional occupations on the individual spatial orbitals.
The electron density is thus given simply as
\begin{equation}
  \label{eq:el_density}
  n_\sigma(r) = \frac{1}{4\pi}\sum_{nl} f^\sigma_{nl} [R^\sigma_{nl}(r)]^2,
\end{equation}
where $f^\sigma_{nl}$ is the number of spin-$\sigma$ electrons on the $nl$ shell.
The occupations $f^\sigma_{nl}$ are determined from the number of electrons with angular momentum $l$ using Hund's rules: the orbitals are filled starting from $n=l+1$ by occupying the $2l+1$ spin-up orbitals for the given $n$ before occupying the corresponding $2l+1$ spin-down orbitals.
We refer to \citerefs{Lehtola2020_PRA_12516} and \citenum{Lehtola2023_JCTC_2502} for further discussion.

The above choice of theory has both a physical and a practical rationale.
The fractional occupation formalism automatically guarantees spherical degeneracy,\cite{Lehtola2020_PRA_12516, Lehtola2023_JCTC_2502} since all magnetic sublevels are assumed to behave identically.
In contrast, DFT calculations with integer occupations of the magnetic sublevels often break spatial symmetry and fail to reproduce the degeneracy of atomic multiplets.\cite{Lehtola2020_PRA_12516}
Although several specialized approaches to reproduce atomic multiplets within a DFT style description have been proposed,\cite{Bagus1975_IJQC_143, Ziegler1977_TCA_261, Barth1979_PRA_1693, Wood1980_JPBAMP_1, Nagy1998_PRA_1672} they appear to be quite involved and have not gained wide adoption.
We also wish to direct the reader to the discussion of \citeit{Baerends1997_CPL_481} on atomic reference energies for DFT calculations.

\subsection{Ionization energies} \label{sec:ion_e}

Within the above level of theory, atoms' ionization energies can be calculated in two ways.
First, in the $\Delta$SCF method the ionization energy is determined for each value of $r_c$ as the difference between the total energy of the cation and of the neutral atom
\begin{equation}
\label{eq:dscf}
E_+(r_c)=E_\mathrm{cation}(r_c)-E_\text{neutral atom}(r_c).
\end{equation}
As is discussed below in \cref{sec:computational-details}, the calculations on unconfined atoms employ the value $r_c=40a_0$ which is large enough to afford fully converged energies for the ground states of atoms and their cations.

Second, \citeit{Boeyens1994_JCSFT_3377} used Koopmans' theorem\cite{Koopmans1934_P_104} to determine the ionization radii.
While there is no Koopmans' theorem for DFT as in Hartree--Fock, Janak's theorem\cite{Janak1978_PRB_7165} provides a justification for an analysis based on orbital energies.
The ionization energy is determined via Janak's theorem as
\begin{equation}
\label{eq:janak}
E_+(r_c)=-\epsilon^\mathrm{HOAO}(r_c),
\end{equation}
where $\epsilon^\mathrm{HOAO}(r_c)$ is the eigenvalue of the highest occupied atomic orbital (HOAO) of the neutral atom with the hard-wall boundary at $r=r_c$.

\subsection{Ionization radii} \label{sec:ion_r_theory}

Now that we have defined ways to calculate ionization energies, we can calculate ionization radii for atoms, i.e., the location of the hard-wall boundary where the atom favors casting away its outermost electron.
The ionization radius $r_+$ for atom X is by solving the nonlinear equation
\begin{equation}
\label{eq:ionization-radius}
E_+(r_c)=0.
\end{equation}

If the ionization energy is computed with \cref{eq:dscf}, \cref{eq:ionization-radius} is straightforward to solve to high precision by bisection, once a crossing between the atomic and cationic curves has been identified.
Now, the issue becomes that the energy of the atom in its neutral and cationic charge state depends on the employed electron configuration.
This issue can be approached in two ways.
The first is to follow the work of \citeit{Boeyens1994_JCSFT_3377} and many others (see \cref{sec:intro}) and fix the electron configuration to that of the ground state of the unconfined neutral atom and its monocation, respectively.
The second is to realize that confinement can affect the ground state configurations of the atom and and its cation, and to compute the energies $E_\text{neutral atom}(r_c)$ and $E_\mathrm{cation}(r_c)$ for the lowest configurations of the neutral atom and cation, respectively, at each value of $r_c$.

Alternatively, we can use the estimate from Janak's theorem of \cref{eq:janak} to solve \cref{eq:ionization-radius}.
Again, one can choose to use the electron configuration of the ground state of the unconfined neutral atom, or to employ a relaxed configuration at every value of $r_c$.
However, exact solution to \cref{eq:ionization-radius} is sometimes problematic in the latter case.
A case in point is the V atom: the value of $r_+$ corresponds to the location of a ground state crossing between a state with $\epsilon^\mathrm{HOAO}(r)<0$ and a state with $\epsilon^\mathrm{HOAO}(r)>0$, which means that \cref{eq:ionization-radius} is not truly satisfied no matter how tightly $r_+$ is converged.

The origin of the above problem is the present use of electron configurations with integer occupations.
However, it has long been known that the use of non-integral occupation numbers for the $4s$ and $3d$ shells can result in a lower total energy for transition metal atoms in average-of-configurations Hartree--Fock calculations.\cite{Slater1969_PR_672}
This phenomenon is also topical for the present DFT calculations,\cite{Kraisler2010_PRA_42516, Lehtola2020_PRA_12516} and has also been found in DFT calculations that do not employ any exchange-correlation functional.\cite{Cances2018_CAMCS_139}
Calculations employing variationally optimized fractional occupation numbers would allow finding an exact solution to \cref{eq:ionization-radius}, likely producing a slightly different radius.
However, we do not consider such fractional occupations in this work, since integer occupation numbers are more commonly used and allow a clearer physical interpretation of the results.

\subsection{Atomic radii \label{sec:atomic-radii}}

The location of the density maximum of the outermost orbital is a long-established estimator for the size of covalently bound atoms\cite{Slater1930_PR_57}
\begin{equation}
  r_\mathrm{max}=\mathrm{max}_i\left(\mathrm{argmax}\left[ r^2n_i(r)\right]\right), \label{eq:densmax}
\end{equation}
where $n_i(r)$ is the $i$th orbital density.

Since the electron density of an atom decays in the far valence
region, a density threshold can be used to estimate the atom's van der
Waals (vdW) radius\cite{Bader1967_JCP_3341}
\begin{equation}
n(r_\rho)=\epsilon, \label{eq:rvdw}
\end{equation}
where the electron density was defined above in \cref{eq:el_density},
and the solution of \cref{eq:rvdw} with a given threshold $\epsilon$
defines the vdW radius estimate $r_\rho$.

Due to several issues related to \cref{eq:rvdw} discovered and
discussed in detail later in this work (for instance, \cref{eq:rvdw} does not
have a solution in some cases, leaving the vdW radius undefined), we
propose another metric for the vdW radius in this work. The vdW radius
can be determined by containing all the electrons of the atom except
$\epsilon$ within the volume enclosed by that radius,
\begin{equation}\label{eq:el_count}
  4\pi\int_{r_\epsilon}^\infty n(r)r^2\mathrm{d}r=\epsilon,
\end{equation}
where the electron density is again defined by \cref{eq:el_density}.
This metric is well defined for any configuration of any element, and
also exhibits superior numerical stability over \cref{eq:rvdw}; see
\cref{sec:radii-results} for discussion. \Cref{eq:el_count} is also
straightforward to extend to non-spherically symmetric electron
densities as
\begin{equation}\label{eq:el_count_generalization}
  \int_{r_\epsilon}^\infty r^2\mathrm{d}r \int n({\bf r}) d\Omega =\epsilon,
\end{equation}
which will be relevant for applications with the CASSCF method, for
instance.

\section{Computational Details} \label{sec:computational-details}

FEM offers a straightforward way to the numerical solution of the radial functions $R^{\sigma}_{nl}$;\cite{Lehtola2019_IJQC_25968} a review of the employed FEM approach can be found in \citerefs{Lehtola2019_IJQC_25945} and \citenum{Lehtola2023_JPCA_4180}.
In short, the radial domain is first divided into $N_\mathrm{elem}$ segments $r \in [r_{i}^\mathrm{start},r_{i}^\mathrm{end}]$ called elements.
An exponential radial grid $\{r_i\}_{i=0}^{N_\mathrm{elem}}$ is employed, so that the size of the elements increases with the distance from the nucleus.\cite{Lehtola2019_IJQC_25945, Lehtola2023_JPCA_4180}
A piecewise polynomial basis of shape functions $B_n(r)$ is then built up in each element.\cite{Lehtola2019_IJQC_25945, Lehtola2023_JPCA_4180}
Finally, the numerical radial basis functions to be used in \cref{eqn:ao} are built from FEM shape functions $B_{n}(r)$ as
\begin{equation}
  \label{eqn:radbas}
    R_{n}(r)=r^{-1}B_n(r),
\end{equation}
and the same radial basis set is used for all angular momenta $l$.
This approach is robust and allows maximal flexibility: as the basis functions within one element have zero overlap to basis functions in other elements, the spatial representation can be adaptively refined, if necessary.

The endpoint of the last element is called the practical infinity, $r_\infty$, and all basis functions are built to vanish at $r_\infty$.\cite{Lehtola2019_IJQC_25945, Lehtola2023_JPCA_4180}
The physical interpretation of $r_\infty$ is that there is a hard-wall potential at this point; \cref{eqn:hard-wall} is therefore already built-in in the FEM approach, and this feature has been previously used in many studies, such as \citerefs{Pasteka2020_MP_1730989} and \citenum{Lehtola2023_JPCA_4180}.
Thus, $r_\infty$ represents a physical parameter that expresses the location of the hard-wall boundary.
In contrast, in studies of unconfined atoms (which represent the typical applications of FEM in the literature), $r_\infty$ is a discretization parameter that needs to be converged such that the obtained solution does not change if an even larger value is employed for $r_\infty$.

As our main focus is on atoms in confinement, the calculations unconfined atoms employed the default value $r_\infty = 40 a_0$, which is sufficient to reproducing the complete basis set limit energy of the ground states of neutral atoms and their cations.
We note here that as larger values of $r_\infty$ may be necessary to capture the behavior of loosely bound states, energies of the excited states in unconfined atoms may not be converged to the complete basis set limit with respect to this parameter.
In contrast, the energies for the excited states in the confined atoms are converged to the complete basis set limit, as is explained below.

All calculations in this work were carried out with the free and open-source\cite{Lehtola2022_WIRCMS_1610} \HelFEM{} program;
the present implementation is publicly available on GitHub.\cite{HelFEM}
The calculations employed shape functions defined by 15-node Lagrange interpolating polynomials (LIPs) specified by Gauss--Lobatto quadrature nodes; this corresponds to the use of a 14$^{\rm th}$ order polynomial basis set.
All calculations were converged to the complete basis set (CBS) limit with respect to the number of radial elements: radial elements were added until a $1 \mu E_h$ precision in the total energy was achieved.
We can thus be assured that we have found the numerically exact wave function.

We employ the optimal FEM implementation of the fractional occupation formalism recently described in \citerefs{Lehtola2020_PRA_12516} and \citenum{Lehtola2023_JCTC_2502}.
We perform calculations within the local density approximation\cite{Bloch1929_ZfuP_545, Dirac1930_MPCPS_376} employing the Perdew--Wang (PW92) correlation functional,\cite{Perdew1992_PRB_13244} within the generalized-gradient approximation (GGA) employing the Perdew--Burke--Ernzerhof (PBE) exchange-correlation functional,\cite{Perdew1996_PRL_3865, Perdew1997_PRL_1396} and within the meta-GGA approximation employing the r$^2$SCAN exchange-correlation functional,\cite{Furness2020_JPCL_8208, Furness2020_JPCL_9248} all as implemented in \Libxc{}\cite{Lehtola2018_S_1} using the \texttt{lda\_x-lda\_c\_pw}, \texttt{gga\_x\_pbe-gga\_c\_pbe}, and \texttt{mgga\_x\_r2scan-mgga\_c\_r2scan} keywords, respectively.

We initiated the work by identifying the three lowest lying configurations of all elements with atomic number $1 \leq Z \leq 54$ in the hard-wall potential for each $r_c\in\{1.0, 1.2, 1.4, \dots, 10.0\}a_0$ with both spin-restricted and spin-polarized densities, respectively, with a fixed angular momentum cutoff $l_\mathrm{max}=3$.
The configurations were determined with the automated algorithm introduced in \citeref{Lehtola2020_PRA_12516}.
The configurations considered for the spin-(un)restricted analysis were then chosen as the union of the low-lying configurations obtained from the spin-(un)restricted search over all of the above values of $r_c$.

We checked whether $g$ orbitals would become occupied for heavy atoms by repeating the configuration search for the Rb and Xe atoms with $l_\mathrm{max}=4$.
However, no configurations with occupied $g$ orbitals were obtained in this search.

The configuration search above was performed with the PBE functional, only.
However, as we shall see in \cref{sec:results}, the results are qualitatively independent of the employed density functional approximation.
For this reason, we are confident that our search yielded all relevant configurations for the PW92 and r$^2$SCAN functionals, as well.

Thus being convinced to have found the low-lying configurations for the H--Xe atoms, we performed calculations on all these configurations with $r_c\in\{1.0,1.1,\dots,10.0\}a_0$.
The wide range of studied values of $r_c$ allows us to observe many interesting evolutions.
The states with $r_c=10.0a_0$ are similar to those of the unconfined atom, while $r_c=1.0a_0$ represents extreme confinement where even core orbitals experience confinement effects and the valence electrons are bound only by the confinement potential.

\section{Results} \label{sec:results}

\subsection{Ionization energies of unconfined atoms} \label{sec:ion_e_2}

As a first step to validate our methods we compute ionization energies for the unconfined H--Xe atoms in the spin-restricted and spin-polarized formalisms according to the two methods discussed in \cref{sec:ion_e}.

\begin{figure*}
\centering
\begin{subfigure}[b]{\textwidth}
\centering
\includegraphics[width=\textwidth]{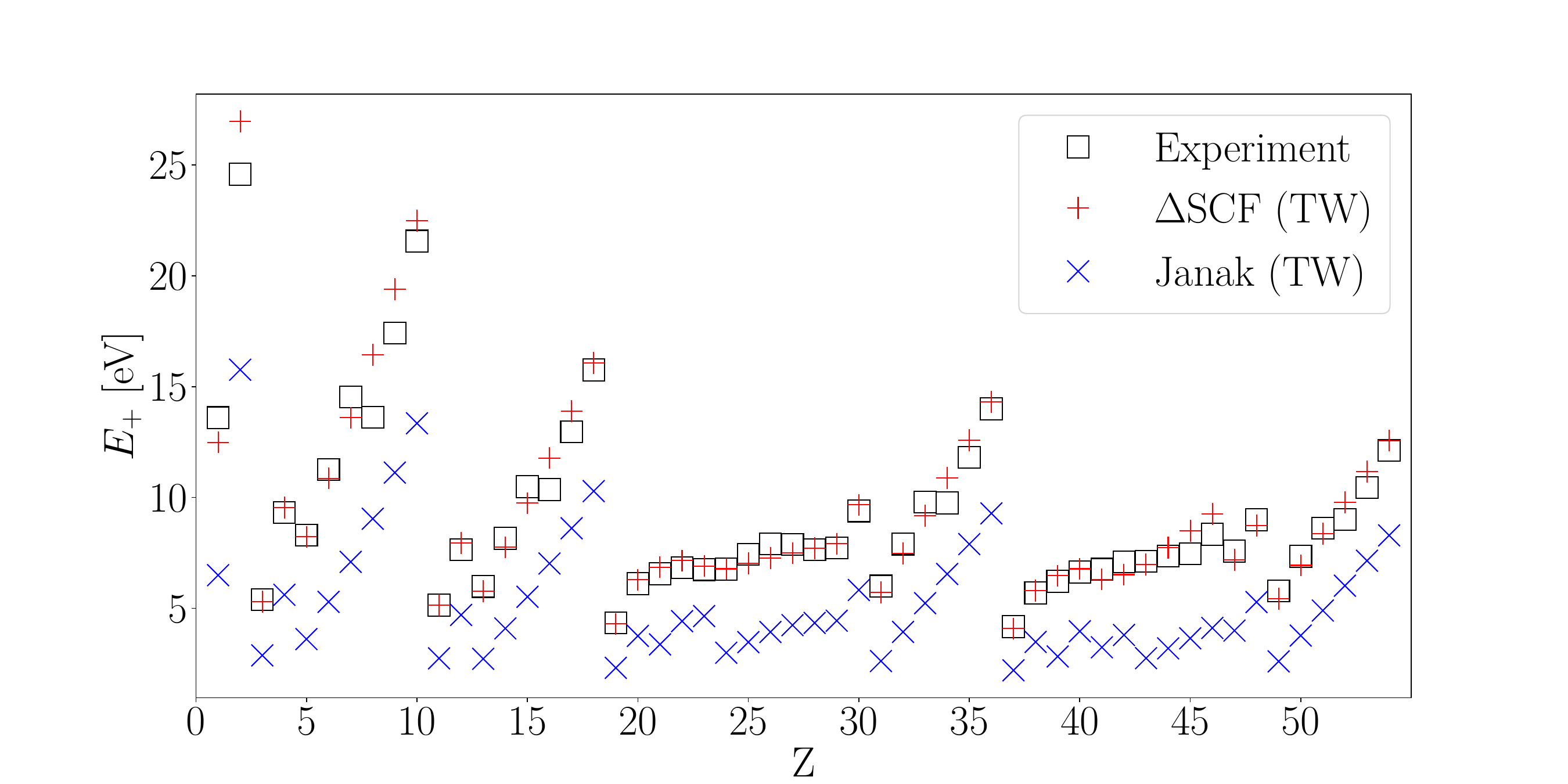}
\caption{Ionization energies of the H--Xe atoms obtained with the PBE functional and spin-restricted densities.}
\label{fig:ion-energies-rest-pbe}
\end{subfigure}
\begin{subfigure}[b]{\textwidth}
\centering
\includegraphics[width=\textwidth]{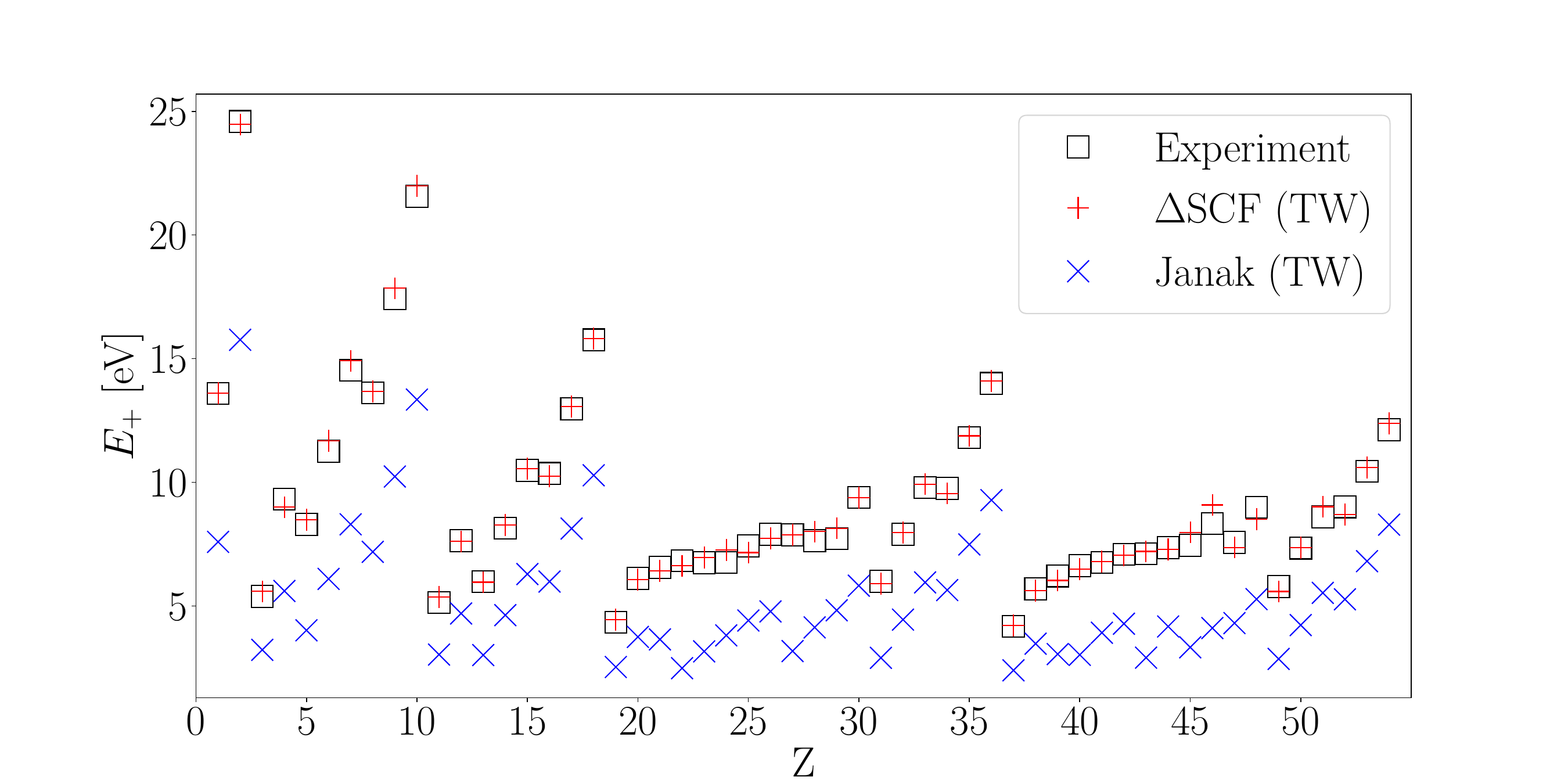}
\caption{Ionization energies of the H--Xe atoms obtained with the PBE functional and spin-polarized densities.}
\label{fig:ion-energies-unrest-pbe}
\end{subfigure}
\caption{Comparison of the ionization energies of unconfined atoms computed with spin-restricted (\cref{fig:ion-energies-rest-pbe}) and spin-polarized (\cref{fig:ion-energies-unrest-pbe}) densities in this work (TW) with $\Delta$SCF or Janak's theorem against experimental values of \citeit{Kramida2023__}.}
\label{fig:ion_e_pbe}
\end{figure*}

As the three dissimilar functionals yield ionization energies that are all in good agreement with the experimental data, we only discuss the PBE data here.
The comparison of our PBE values to the experimental values of \citeit{Kramida2023__} is depicted in \cref{fig:ion_e_pbe}; analogous plots for the PW92 and r$^2$SCAN functionals are included in the SI.

The values obtained from Janak's theorem are consistently and significantly lower than the experimental values and we will therefore not discuss them further.

The $\Delta$SCF formalism produces ionization energies in good agreement with experiment.
As expected, the results of the spin-polarized calculations are in better agreement with experiment than those from the spin-restricted calculations.
Notable differences can be observed already for the H and He atoms.
Spin-polarized calculations also faithfully reproduce the dip in the ionization energy beyond half filling of the $p$ shell, while spin-restricted calculations predict that the ionization energy increases monotonically with the filling of the $p$ shell.

\subsection{Ionization radii} \label{sec:ion_radii}

\subsubsection{$\Delta$SCF \label{sec:deltascf}}

\begin{sidewaystable}
\centering
\begin{tabular}{cccccccccccccccccc}
\makecell{ H \\ \color{red}{ 1.95 } \\ \color{blue}{ 1.85 } } & &\makecell{ Ca \\ \color{red}{ 4.60 } \\ \color{blue}{ 4.66 } } &\makecell{ Sc \\ \color{red}{ 4.37 } \\ \color{blue}{ 4.46 } } &\makecell{ Ti \\ \color{red}{ 4.53 } \\ \color{blue}{ 4.27 } } &\makecell{ V \\ \color{red}{ 4.43 } \\ \color{blue}{ 4.09 } } &\makecell{ Cr \\ \color{red}{ 4.08 } \\ \color{blue}{ 3.94 } } &\makecell{ Mn \\ \color{red}{ 3.94 } \\ \color{blue}{ 3.95 } } &\makecell{ Fe \\ \color{red}{ 3.83 } \\ \color{blue}{ 3.79 } } &\makecell{ Co \\ \color{red}{ 3.72 } \\ \color{blue}{ 3.65 } } &\makecell{ Ni \\ \color{red}{ 3.63 } \\ \color{blue}{ 3.57 } } & & & & & & & \makecell{ He \\ \color{red}{ 1.34 } \\ \color{blue}{ 1.40 } } \\ 
\makecell{ Li \\ \color{red}{ 4.17 } \\ \color{blue}{ 4.06 } } &\makecell{ Be \\ \color{red}{ 3.14 } \\ \color{blue}{ 3.21 } } &\makecell{ Sr \\ \color{red}{ 4.97 } \\ \color{blue}{ 5.02 } } &\makecell{ Y \\ \color{red}{ 4.72 } \\ \color{blue}{ 4.82 } } &\makecell{ Zr \\ \color{red}{ 4.87 } \\ \color{blue}{ 4.55 } } &\makecell{ Nb \\ \color{red}{ 4.52 } \\ \color{blue}{ 4.37 } } &\makecell{ Mo \\ \color{red}{ 4.37 } \\ \color{blue}{ 4.22 } } &\makecell{ Tc \\ \color{red}{ 3.58 } \\ \color{blue}{ 4.12 } } &\makecell{ Ru \\ \color{red}{ 3.39 } \\ \color{blue}{ 4.04 } } & & & & \makecell{ B \\ \color{red}{ 3.05 } \\ \color{blue}{ 3.01 } } & \makecell{ C \\ \color{red}{ 2.66 } \\ \color{blue}{ 2.58 } } & \makecell{ N \\ \color{red}{ 2.37 } \\ \color{blue}{ 2.28 } } & \makecell{ O \\ \color{red}{ 2.15 } \\ \color{blue}{ 2.27 } } & \makecell{ F \\ \color{red}{ 1.96 } \\ \color{blue}{ 2.01 } } & \makecell{ Ne \\ \color{red}{ 1.81 } \\ \color{blue}{ 1.83 } } \\ 
\makecell{ Na \\ \color{red}{ 4.38 } \\ \color{blue}{ 4.30 } } &\makecell{ Mg \\ \color{red}{ 3.70 } \\ \color{blue}{ 3.76 } } & & & & & & & & & & & \makecell{ Al \\ \color{red}{ 4.13 } \\ \color{blue}{ 4.08 } } & \makecell{ Si \\ \color{red}{ 3.64 } \\ \color{blue}{ 3.55 } } & \makecell{ P \\ \color{red}{ 3.28 } \\ \color{blue}{ 3.18 } } & \makecell{ S \\ \color{red}{ 2.99 } \\ \color{blue}{ 3.11 } } & \makecell{ Cl \\ \color{red}{ 2.75 } \\ \color{blue}{ 2.80 } } & \makecell{ Ar \\ \color{red}{ 2.56 } \\ \color{blue}{ 2.57 } } \\ 
\makecell{ K \\ \color{red}{ 5.26 } \\ \color{blue}{ 5.18 } } &\makecell{ Ca \\ \color{red}{ 4.82 } \\ \color{blue}{ 4.77 } } &\makecell{ Sc \\ \color{red}{ 4.62 } \\ \color{blue}{ 4.49 } } &\makecell{ Ti \\ \color{red}{ 4.40 } \\ \color{blue}{ 4.14 } } &\makecell{ V \\ \color{red}{ 4.13 } \\ \color{blue}{ 3.82 } } &\makecell{ Cr \\ \color{red}{ 3.88 } \\ \color{blue}{ 3.94 } } &\makecell{ Mn \\ \color{red}{ 3.67 } \\ \color{blue}{ 3.83 } } &\makecell{ Fe \\ \color{red}{ 3.50 } \\ \color{blue}{ 3.72 } } &\makecell{ Co \\ \color{red}{ 3.35 } \\ \color{blue}{ 3.46 } } &\makecell{ Ni \\ \color{red}{ 3.22 } \\ \color{blue}{ 3.25 } } &\makecell{ Cu \\ \color{red}{ 3.54 } \\ \color{blue}{ 3.50 } } &\makecell{ Zn \\ \color{red}{ 3.30 } \\ \color{blue}{ 3.33 } } &\makecell{ Ga \\ \color{red}{ 4.06 } \\ \color{blue}{ 4.01 } } &\makecell{ Ge \\ \color{red}{ 3.71 } \\ \color{blue}{ 3.62 } } &\makecell{ As \\ \color{red}{ 3.44 } \\ \color{blue}{ 3.34 } } &\makecell{ Se \\ \color{red}{ 3.21 } \\ \color{blue}{ 3.33 } } &\makecell{ Br \\ \color{red}{ 3.02 } \\ \color{blue}{ 3.07 } } &\makecell{ Kr \\ \color{red}{ 2.86 } \\ \color{blue}{ 2.87 } } \\ 
\makecell{ Rb \\ \color{red}{ 5.56 } \\ \color{blue}{ 5.49 } } &\makecell{ Sr \\ \color{red}{ 5.12 } \\ \color{blue}{ 5.07 } } &\makecell{ Y \\ \color{red}{ 4.85 } \\ \color{blue}{ 4.59 } } &\makecell{ Zr \\ \color{red}{ 4.41 } \\ \color{blue}{ 4.15 } } &\makecell{ Nb \\ \color{red}{ 4.07 } \\ \color{blue}{ 3.83 } } &\makecell{ Mo \\ \color{red}{ 3.80 } \\ \color{blue}{ 4.17 } } &\makecell{ Tc \\ \color{red}{ 3.58 } \\ \color{blue}{ 3.79 } } &\makecell{ Ru \\ \color{red}{ 3.39 } \\ \color{blue}{ 3.51 } } &\makecell{ Rh \\ \color{red}{ 3.23 } \\ \color{blue}{ 3.29 } } &\makecell{ Pd \\ \color{red}{ 3.09 } \\ \color{blue}{ 3.10 } } &\makecell{ Ag \\ \color{red}{ 3.88 } \\ \color{blue}{ 3.84 } } &\makecell{ Cd \\ \color{red}{ 3.66 } \\ \color{blue}{ 3.69 } } &\makecell{ In \\ \color{red}{ 4.41 } \\ \color{blue}{ 4.37 } } &\makecell{ Sn \\ \color{red}{ 4.09 } \\ \color{blue}{ 4.00 } } &\makecell{ Sb \\ \color{red}{ 3.83 } \\ \color{blue}{ 3.73 } } &\makecell{ Te \\ \color{red}{ 3.62 } \\ \color{blue}{ 3.73 } } &\makecell{ I \\ \color{red}{ 3.43 } \\ \color{blue}{ 3.48 } } &\makecell{ Xe \\ \color{red}{ 3.27 } \\ \color{blue}{ 3.28 } } \\ 
\end{tabular}
\caption{\small{Ionization radii $R$ in $a_0$ for the H--Xe atoms from spin-restricted (\textcolor{red}{in red}) and spin-polarized (\textcolor{blue}{in blue}) non-relativistic $\Delta$SCF calculations with fractional occupations and the PBE functional with configurations fixed to the ground states of the unconfined neutral atom and its cation. Values obtained while considering the $R$ dependence of the ground state of the neutral atom and its cation are highlighted in the insert for the elements where the value differs from the static configurations.}}
\label{tab:ion_r_pbe}
\end{sidewaystable}

As discussed in \cref{sec:ion_r_theory}, $\Delta$SCF ionization radii can be computed in two ways.
In the first method we fix the configurations to those of the ground state of the neutral atoms and their cations, respectively.
In the second method we relax the configurations of the atoms and cations to the lowest configurations at each radius.
The resulting ionization radii for the H--Xe atoms obtained with the PBE functional are shown in \cref{tab:ion_r_pbe} and its insert, respectively, when the second method predicts a different ionization radius than the fixed configuration approach.

The ionization radius for the hydrogen atom is an interesting place to start the analysis.
The corresponding exact model is analytically solvable: \citeit{Sommerfeld1938_APB_56} obtained the ionization radius $1.8352a_0$ in their pioneering study.
The exact ground state of hydrogen is fully spin polarized and our spin-polarized PW92, PBE and r$^2$SCAN values of $1.90a_0$, $1.86a_0$, and $1.84a_0$ deviate from the exact value by only 4.5\%, 1.4\% and 0.3\%, respectively.
The spin-restricted values of $1.99a_0$, $1.95a_0$, and $1.96a_0$ exhibit much larger deviations of 8.4\%, 6.3\%, and 6.8 \%, respectively, but are still remarkably accurate considering that our spin-restricted calculations are for an atom with $1/2$ spin-up and $1/2$ spin-down electrons.

Moving on to heavier elements, the comparison of the results from the fixed electron configuration and relaxed configuration methods proves interesting.
Although inclusion of the relaxation effects does not appear to be important for most of the studied elements, large differences are observed in the pre-$d$ and $d$ blocks of the periodic table (Ca--Ni and Sr--Ru).

Since relaxing the configuration can result in a considerable decrease of the energy of the neutral atom, the second method is na\"ively expected to yield ionization radii that are considerably smaller than the radii obtained with the first method of examining fixed electron configurations for the neutral atom and its cation.
As expected, allowing the configuration to relax results in significant decreases in the ionization radii for most transition metals: for instance, the ionization radius of Mo decreases by 13\% in the spin-restricted calculations when changes to the ground state of Mo and \ce{Mo+} are considered (the spin-polarized change is just $-1.4\%$).
Similarly, the spin-polarized calculation on Ru leads to an ionization radius that is $13\%$ smaller when the electron configuration is allowed to relax (no change in the spin-restricted calculation).

\begin{figure*}
\centering
\begin{subfigure}[b]{0.48\textwidth}
\centering
\includegraphics[width=\textwidth]{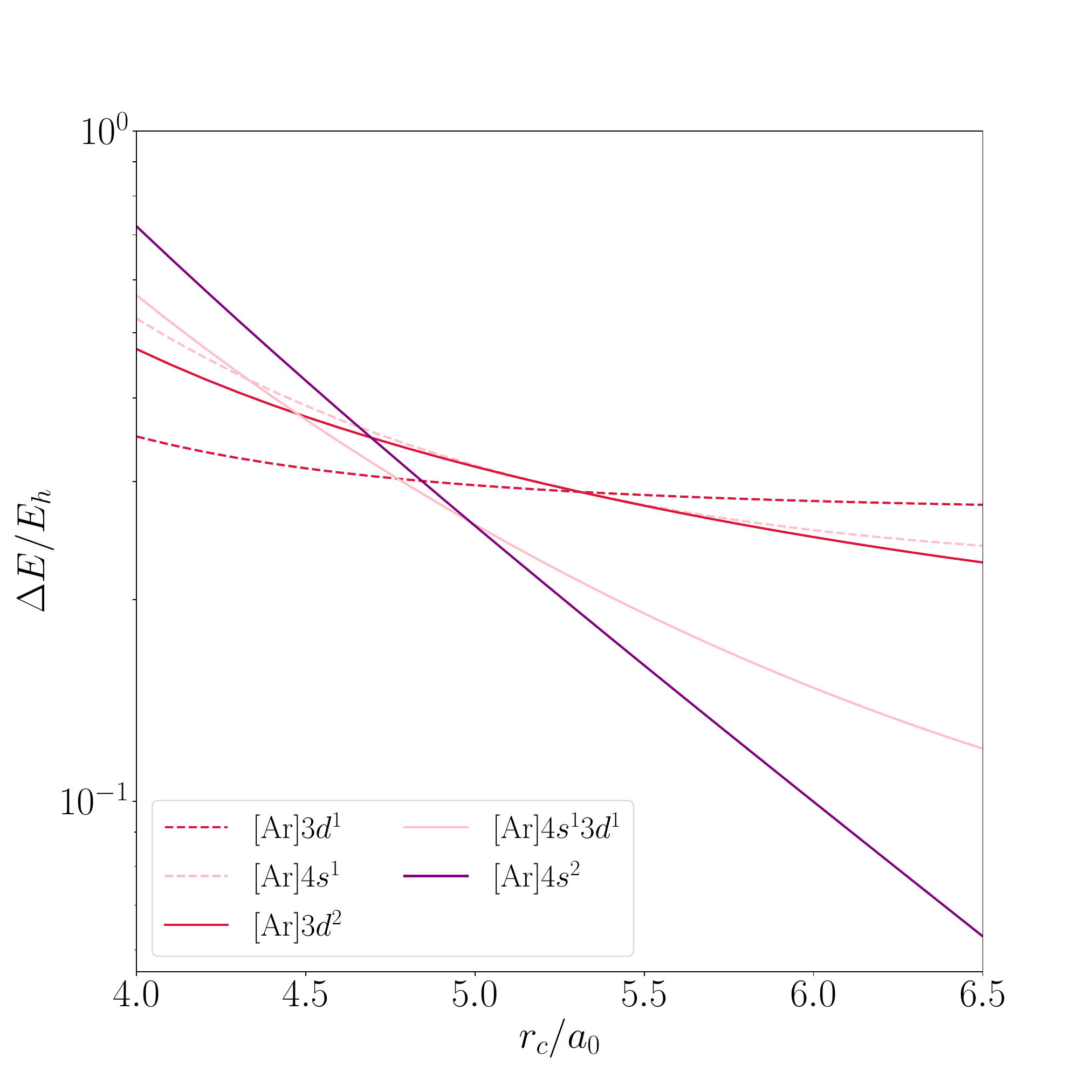}
\caption{Ca}
\end{subfigure}
\begin{subfigure}[b]{0.48\textwidth}
\centering
\includegraphics[width=\textwidth]{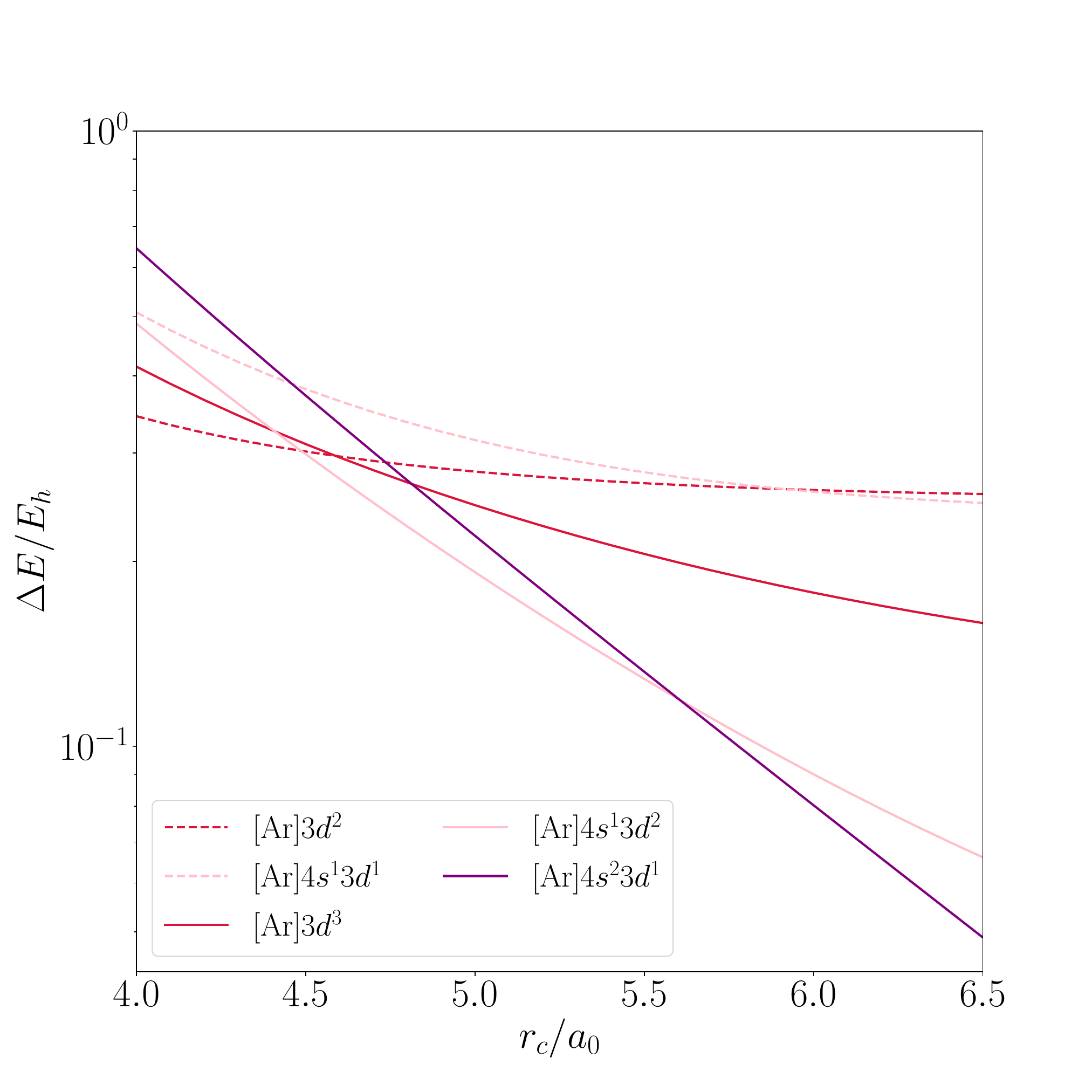}
\caption{Sc}
\end{subfigure}
\begin{subfigure}[b]{0.48\textwidth}
\centering
\includegraphics[width=\textwidth]{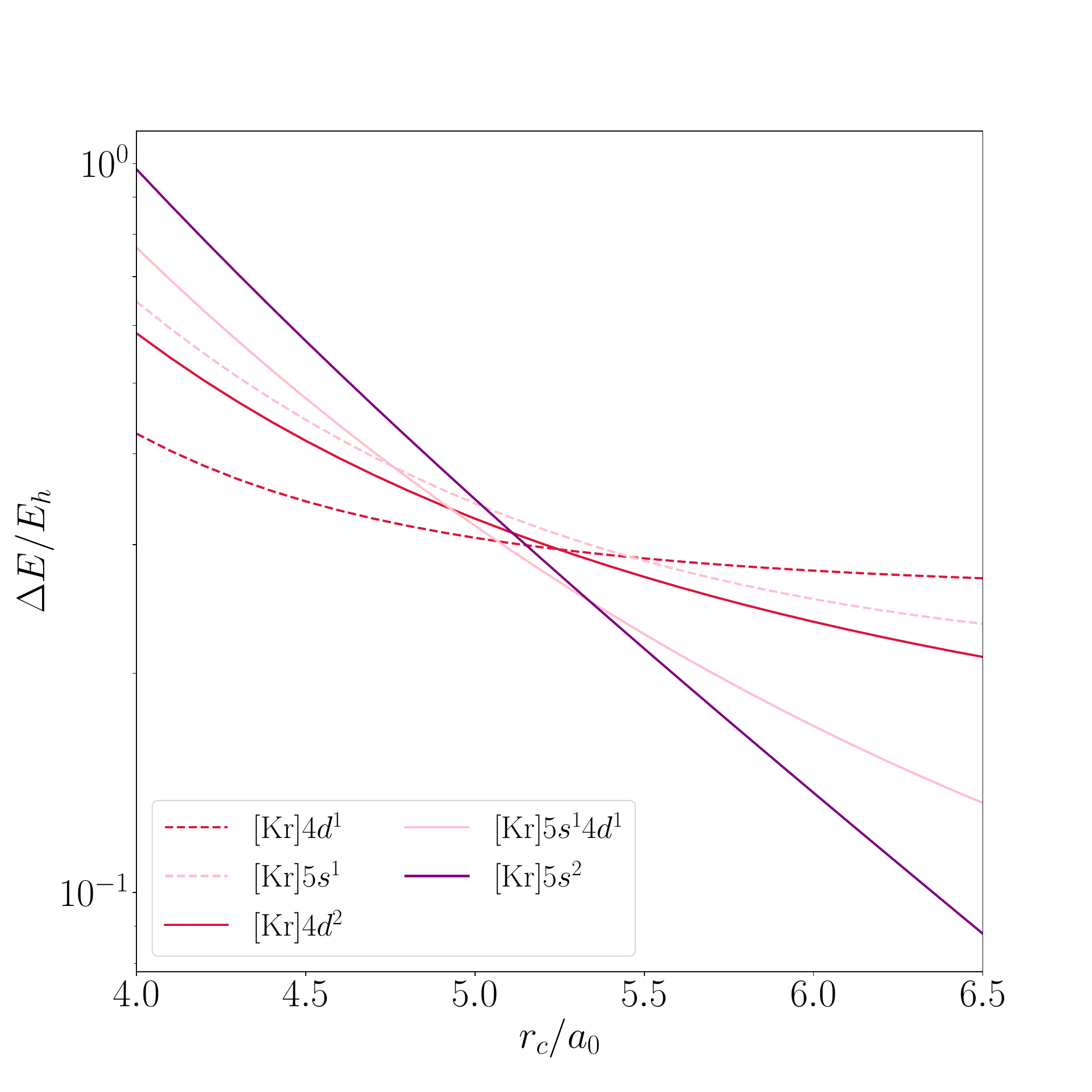}
\caption{Sr}
\end{subfigure}
\begin{subfigure}[b]{0.48\textwidth}
\centering
\includegraphics[width=\textwidth]{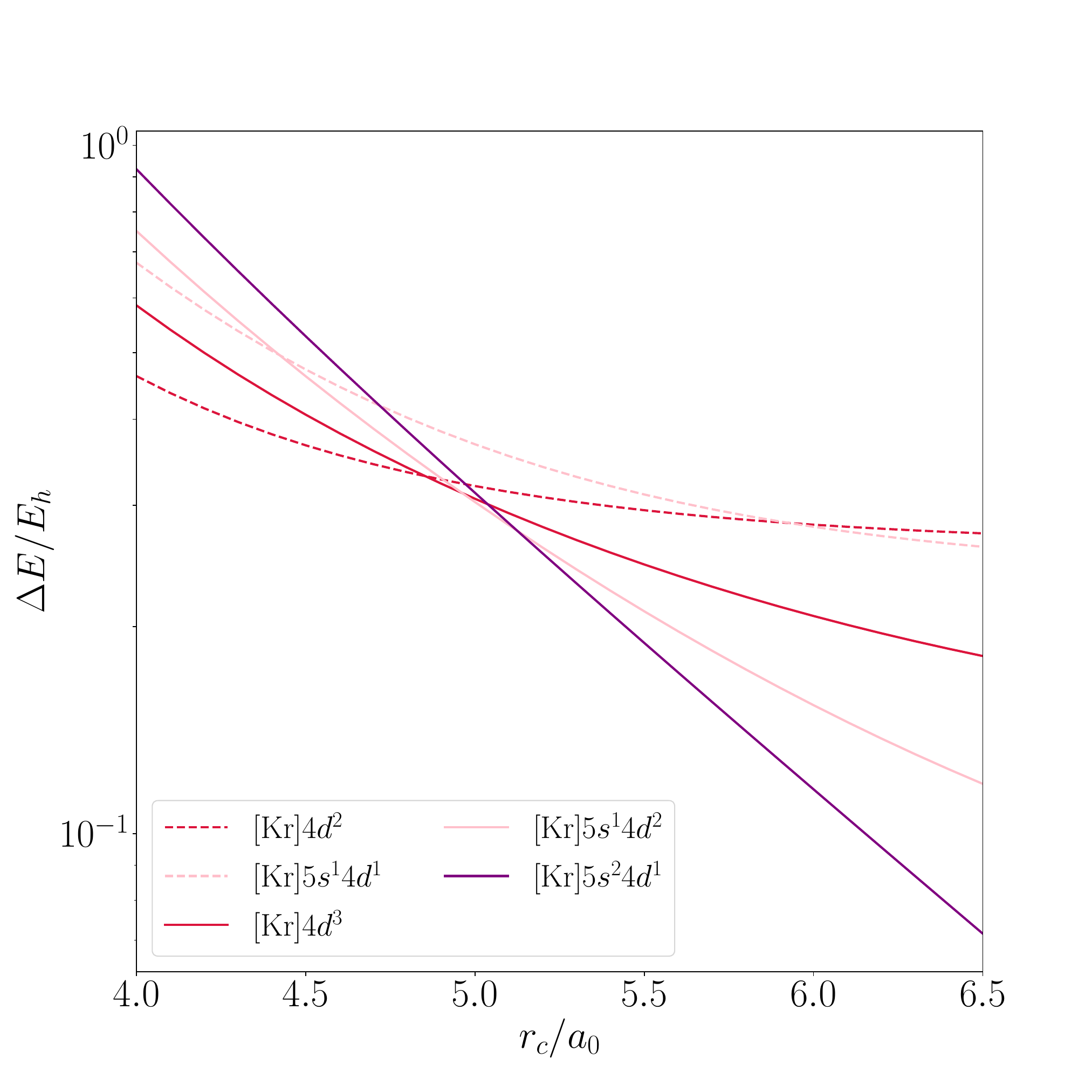}
\caption{Y}
\end{subfigure}
\caption{Ground state crossings for the spin-polarized Ca, Sc and Sr atoms, and the spin-restricted Y atom in the hard-wall potential with the PBE functional. Configurations for the neutral atoms shown as solid lines, and those corresponding to the monocation as dashed lines.}
\label{fig:ion_r_case-study}
\end{figure*}

However, significant changes in energy upon relaxing the configuration may also occur for the cation, in which case the ionization radius can increase.
Interestingly, such increases when the ground states are relaxed \emph{are} observed for the Ca, Sc, and Sr atoms in both the spin-restricted and spin-polarized cases, as well as the Y atom in the spin-restricted case.
The corresponding cations exhibit a ground state crossing that moves the cross-over point of \cref{eq:ionization-radius} down in energy, and thereby the ionization radius to the right; see \cref{fig:ion_r_case-study} for an illustration of these four exceptional cases.

Because the various configurations' energies span several orders of magnitude under the studied range of confinement radii, we compare the energy of the confined atom with electron configuration $i$ to that of the ground state configuration of the unconfined atom
\begin{equation}
\Delta E_i(r_c)=E_i(r_c)-E_0^\mathrm{unconfined},
\end{equation}
where $i$ is the state in question in all plots in this study.

The differences observed in ionization radii with and without relaxing the electron configuration underline the importance of considering all possible low lying configurations as a function of confinement.
Many transition metal atoms and their monocations exhibit ground state crossings already in weak confinement, due to the low-lying $s$-$d$ excitations in these systems (see \cref{sec:changes}).

The spin-polarized and spin-restricted ionization radii differ for all atoms, which is not surprising given that either the neutral atom or its cation is going to be open-shell.
Interestingly, the differences between these two approaches are the smallest for the noble gas atoms, while alkali metals show much larger differences between spin-polarized and spin-unpolarized calculations.
As the spin-polarized calculations gave ionization energies in best agreement with experiment in \cref{sec:ion_e_2}, the spin-polarized ionization radii should be the most accurate.

\begin{figure*}
\centering
\includegraphics[width=\textwidth]{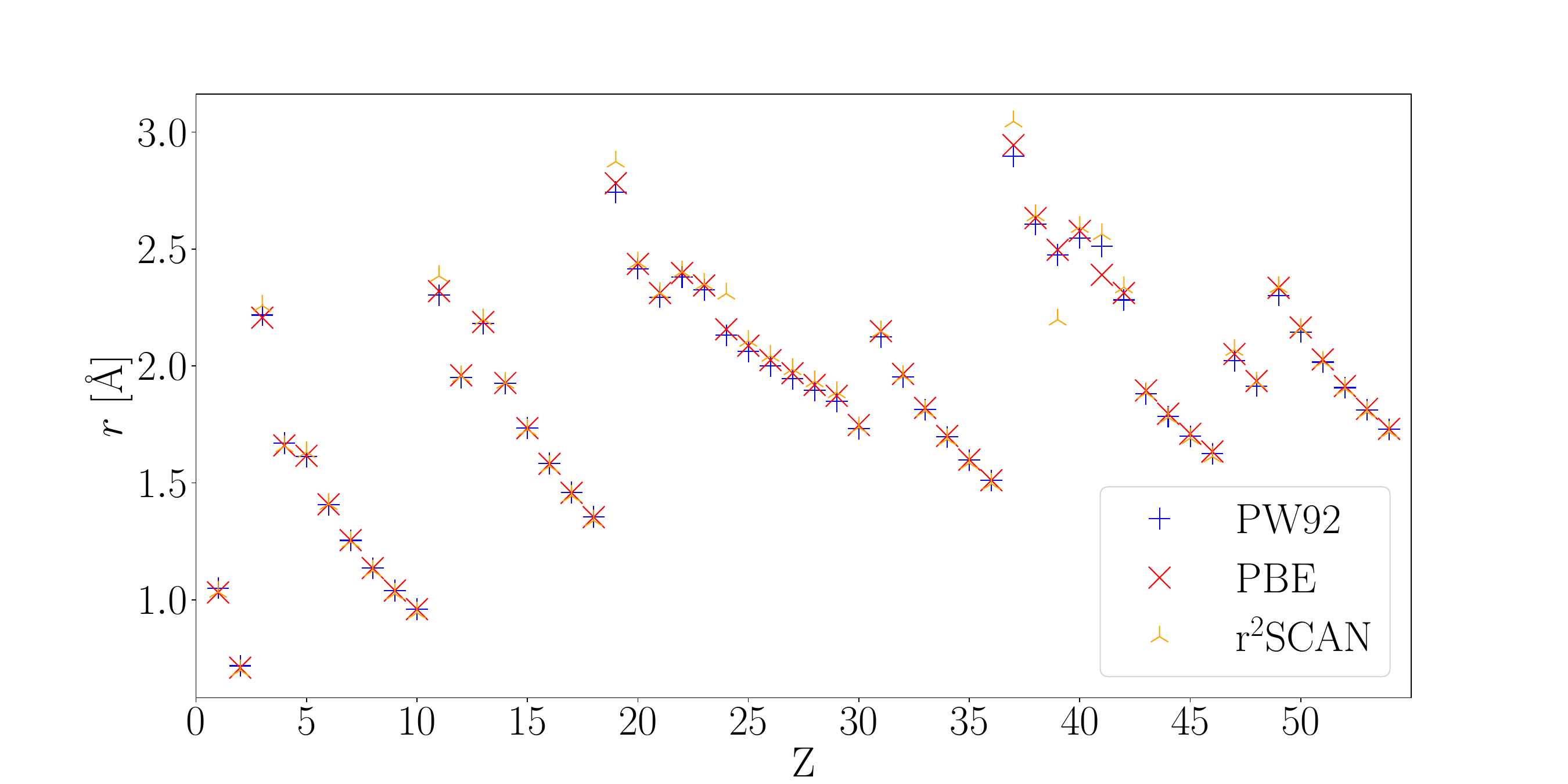}
\caption{Ionization radii for the H--Xe atoms with the PW92, PBE and r$^2$SCAN functionals obtained from spin-polarized densities via $\Delta$SCF with relaxed configurations.}
\label{fig:compare_ion_r}
\end{figure*}

\begin{figure}
\centering
\includegraphics[width=0.48\textwidth]{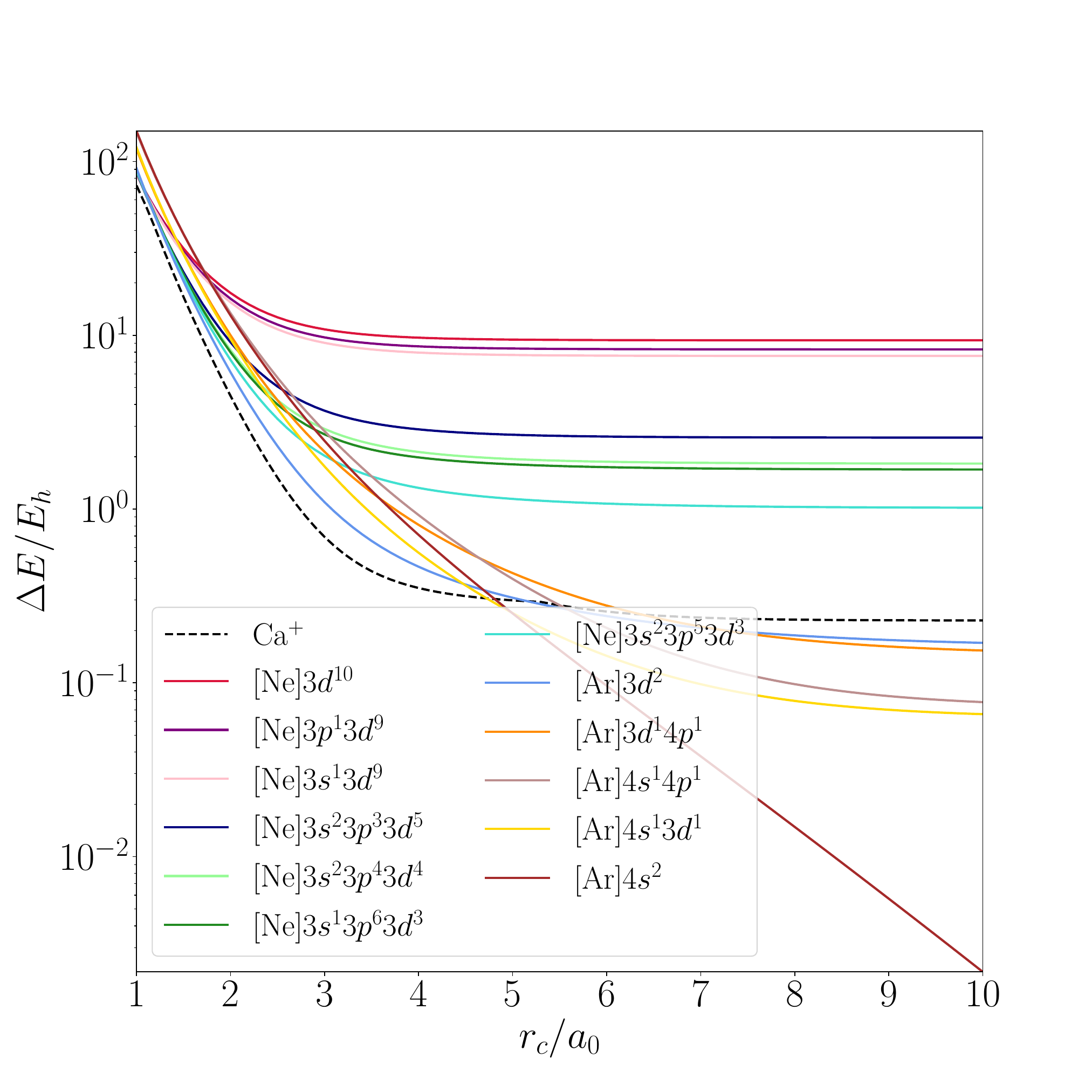}
\caption{Energies of low lying configurations of hard-wall confined spin-polarized Ca computed with PW92 shown as the energy difference from unconfined Ca as a function of the confinement radius. Note semilogarithmic scale.}
\label{fig:ca-pbe}
\end{figure}

\begin{figure}
\centering
\includegraphics[width=0.48\textwidth]{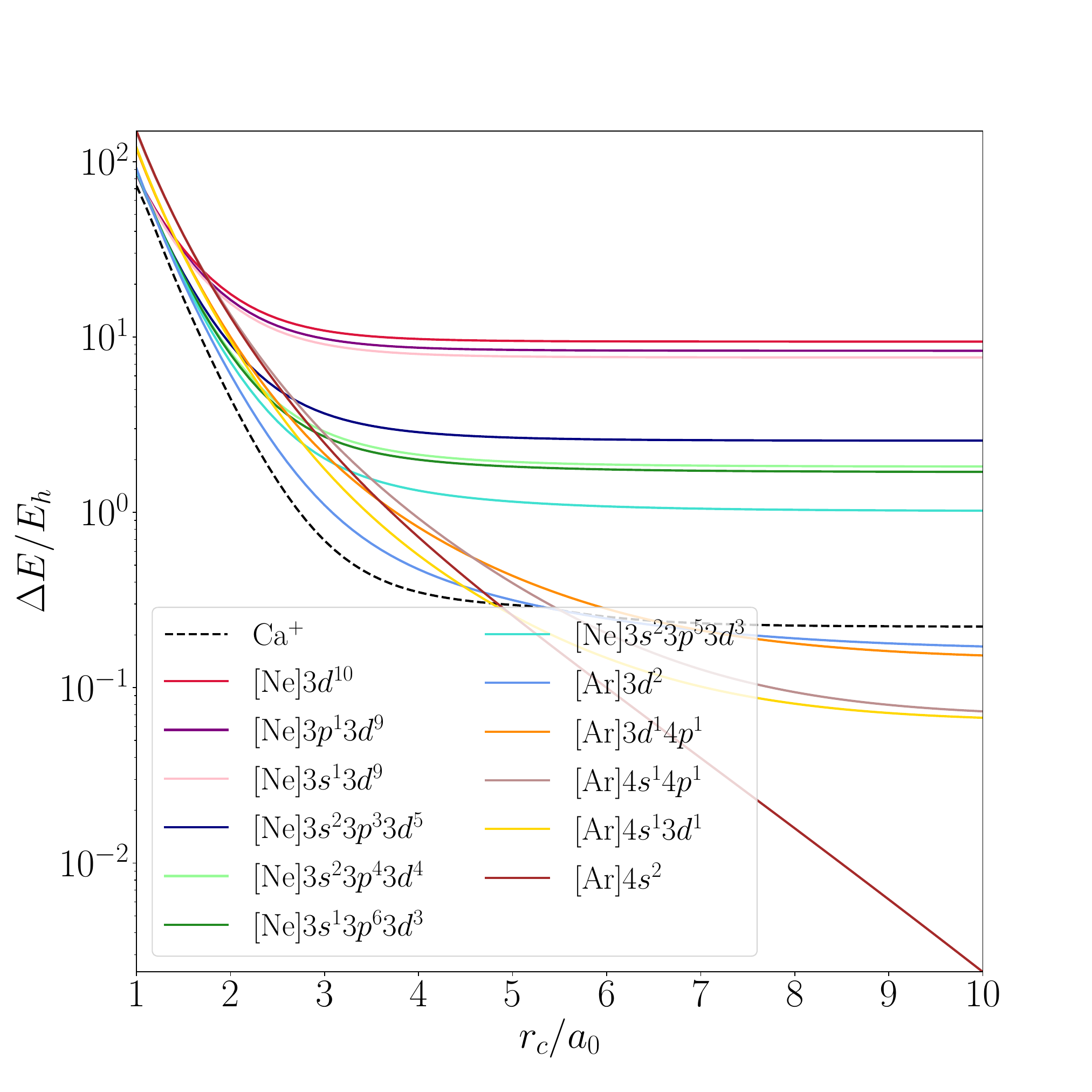}
\caption{Energies of low lying configurations of hard-wall confined spin-polarized Ca computed with PBE shown as the energy difference from unconfined Ca as a function of the confinement radius. Note semilogarithmic scale.}
\label{fig:ca-pw}
\end{figure}

\begin{figure}
\centering
\includegraphics[width=0.48\textwidth]{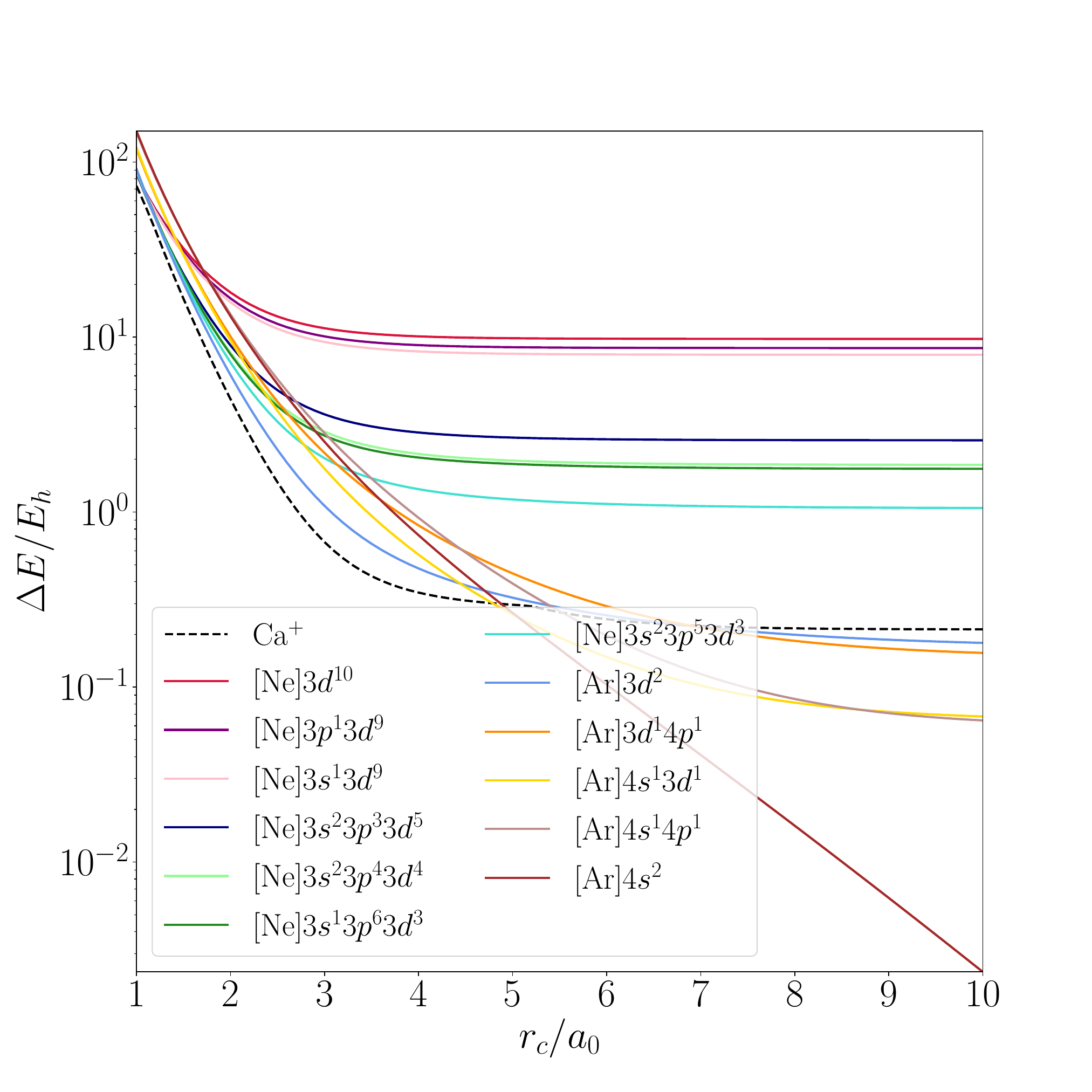}
\caption{Energies of low lying configurations of hard-wall confined spin-polarized Ca computed with r$^2$SCAN shown as the energy difference from unconfined Ca as a function of the confinement radius. Note semilogarithmic scale.}
\label{fig:ca-r2scan}
\end{figure}

A comparison of the ionization radii obtained with the PW92, PBE and r$^2$SCAN functionals with spin-polarized densities and relaxed configurations is shown in \cref{fig:compare_ion_r}.
Except for the alkali metals, as well as the Cr, Y, and Nb atoms, the dissimilar density functional approximations are in excellent agreement.
We further validate this agreement by qualitatively studying the functional dependence of the behavior of all low lying configurations of the Ca atom in \cref{fig:ca-pw,fig:ca-pbe,fig:ca-r2scan}, while the graphical representation of the results of the other atoms is included in the SI.
We chose to feature Ca here due to its interesting chemistry which will be discussed below in \cref{sec:changes}.
The only major difference between the functionals is that r$^2$SCAN predicts the $4s^1 3d^1$ configuration to be lower in energy than the $4s^1 4p^1$ configuration in the unconfined atom, while PW92 and PBE predict the opposite.

Despite such minor differences in the ordering of some states, or relative energy differences between states, the qualitative behaviors obtained with the three different density functional approximations agree well with each other; the changes in the electron configurations are practically independent of the functional.
This suggests that the phenomena found with the methodology employed in this work are not artefacts of the used level of theory, but true physical phenomena within the studied model of confinement.

\subsubsection{Janak's theorem \label{sec:janak}}

We report ionization radii obtained from calculations using Janak's theorem as tables in the SI, as we already found above in \cref{sec:ion_e_2} that these calculations failed to reproduce ionization energies in a reliable manner.

\citeit{Sen2000_CPL_29} computed atomic ionization radii with Janak's theorem for the He--Ne atoms with exchange-only calculations based on Becke's 1988 generalized-gradient functional.\cite{Becke1988_PRA_3098}
In contrast to our work, \citeit{Sen2000_CPL_29} employed the transition state approximation (TSA) of \citeit{Slater1969_PR_672} in their calculations; the orbital energies were thus determined in calculations where half an electron was removed from the highest-lying orbital.
\citeit{Sen2000_CPL_29} compared this method to $\Delta$SCF calculations and found that the two methods yield ionization radii in close agreement to each other.
Despite the differences between our approaches, we will see below that our results are in good agreement with those of \citeit{Sen2000_CPL_29}.

\subsubsection{Comparison to literature radii \label{sec:comparison}}

\begin{figure*}
\centering
\includegraphics[width=\textwidth]{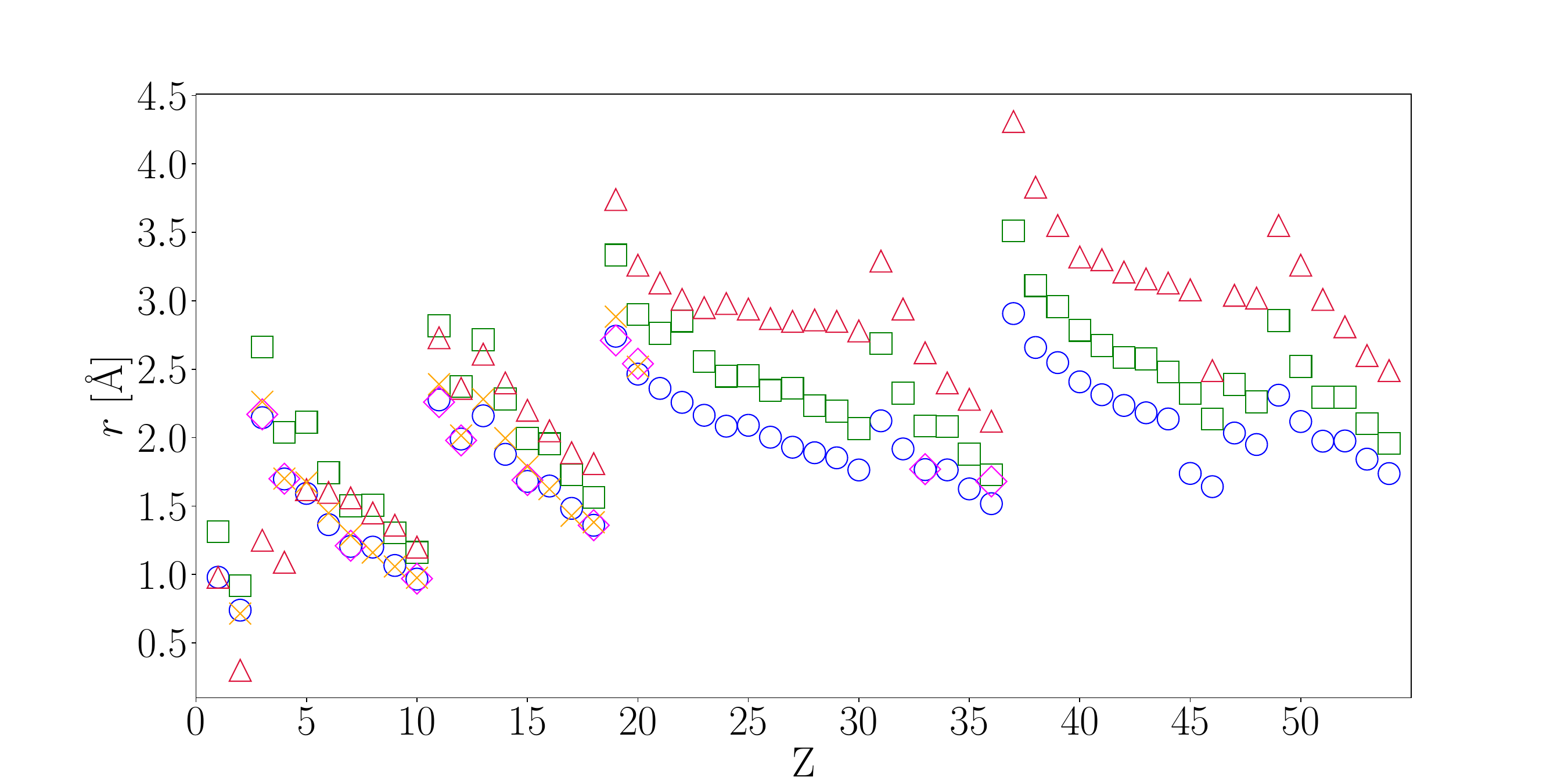}
\caption{Comparison of ionization radii computed in this work with spin-polarized densities and the PBE density functional via $\Delta$SCF (blue circles) or Janak's theorem (green squares) against the Hartree--Fock values of \citeit{{Boeyens1994_JCSFT_3377}} (red triangles) and DFT values of \citeit{{Sen2000_CPL_29}} (magenta diamonds) and \citeit{{Garza2005_JCS_379}} (orange crosses).}
\label{fig:r_comp_False_pbe}
\end{figure*}

Numerically reliable ionization radii have been previously published by \citeit{Boeyens1994_JCSFT_3377}, \citeit{Sen2000_CPL_29}, and \citeit{Garza2005_JCS_379}.
We compare our ionization radii from calculations based on the $\Delta$SCF method (\cref{sec:deltascf}) and Janak's theorem (\cref{sec:janak}) with relaxed configurations and the PBE functional against the earlier calculations in \cref{fig:r_comp_False_pbe}.
Analogous plots for the PW92 and r$^2$SCAN functionals can be found in the SI.
We note that the dip in energy for the group 16 elements, which was found in the ionization energies in \cref{sec:ion_e_2}, is also visible in the ionization radii in \cref{fig:r_comp_False_pbe}.

We first note that  the analysis of \citeit{Boeyens1994_JCSFT_3377} was based on non-relativistic Hartree--Fock calculations for $1 \leq Z \leq 102$, and $r_+$ was identified via Koopmans' theorem\cite{Koopmans1934_P_104} by finding the zero of the orbital energy of the outermost valence electron shell.
However, as discussed in \cref{sec:intro}, \citeit{Boeyens1994_JCSFT_3377} did not employ a proper hard-wall potential, and did not consider the changing nature of the ground state configuration as a function of confinement.
The values of \citeit{Boeyens1994_JCSFT_3377} are systematically and considerably larger than the results of any of the other calculations for $Z\ge 15$.
These surprisingly large differences are likely explained by the lack of electron correlation in Hartree--Fock theory, which is known to lead to overestimation of atoms' size compared to high-level wave function calculations, with DFT yielding close agreement with accurate theoretical reference values.\cite{Cohen2004_PCCP_2928}
The closeness of DFT densities with multideterminantal densities has also been noted on previously by \citeit{OrtizHenarejos1997_IJQC_245}, for example.

Our calculations based on Janak's theorem without the TSA systematically predict larger ionization radii than our $\Delta$SCF calculations, but they are still much smaller than the Hartree--Fock values of \citeit{Boeyens1994_JCSFT_3377}.
Our $\Delta$SCF calculations are in excellent agreement with the limited data reported by \citeit{Sen2000_CPL_29} in the spin-restricted formalism, who also found good agreement between their TSA method and $\Delta$SCF calculations in their work.
Interestingly, our spin-polarized data appears to be in better agreement with the calculations of \citeit{Sen2000_CPL_29} than our spin-restricted calculations, as the additional comparisons included in the SI show, even though the calculations of \citeit{Sen2000_CPL_29} were spin-restricted.
As was discussed in \cref{sec:janak}, \citeit{Sen2000_CPL_29} did not include correlation in their calculations, which likely explains the differences between our data.

Our $\Delta$SCF calculations are also in good agreement for the 9 atoms studied by \citeit{Garza2005_JCS_379} with spin-polarized $\Delta$SCF calculations with the PW92 functional.

Since our $\Delta$SCF calculations were found to produce ionization potentials of unconfined atoms in good agreement with experimental values (\cref{sec:ion_e_2}), and also yield ionization radii in good agreement between dissimilar density functional approximations as well as with previous literature, we are confident that our calculations offer thus a reliable basis for studies of confined atoms, and a reliable ground for further studies with elaborate wave function methods.

\subsection{Estimates for atomic radii \label{sec:radii-results}}

Before proceeding to the in-depth analysis of the changes of atoms'
electronic structure in confinement, we return to the question of the
atomic radii discussed above in \cref{sec:atomic-radii}.  Because the
present study involves many configurations for each atom, many not
being bound with respect to ionization in the unconfined atom, it is
useful to be able to estimate the resulting atoms' size.  While the
covalent size estimate of \citeit{Slater1930_PR_57} in
\cref{eq:densmax} is parameter-free, the two vdW radius estimates of
\cref{eq:rvdw,eq:el_count} feature a threshold parameter.

\citeit{Bader1967_JCP_3341} employed the threshold 0.002
electrons/bohr$^3$ in \cref{eq:rvdw}, while
\citeit{Boyd1977_JPBAMP_2283} showed that the criterion 0.001
electrons/bohr$^3$ reproduces the same relative radii of atoms as
those obtained with an even smaller value for the threshold.
\citeauthor{Rahm2016_CEJ_14625} also employed the threshold 0.001
electrons/bohr$^3$ in their study of atomic and ionic
radii\cite{Rahm2016_CEJ_14625, Rahm2017_CEJ_4017} and we chose to
adopt this value of the threshold, $\epsilon = 10^{-3}$ in
\cref{eq:rvdw}, as well.  We note that \citeit{Smith2023_JCTC_2064}
have also used \cref{eq:rvdw} recently to estimate vdW
radii, opting to use a threshold of 0.0015 electrons/bohr$^3$,
instead.

We discovered an issue in our analysis with the $r_\rho$ estimate of
\cref{eq:rvdw} for the vdW radius for the $3d^1$ state of the H atom.
As only $s$-type orbitals can have a finite value at the
nucleus,\cite{Lehtola2019_IJQC_25945} and as the $3d$ orbital in
question turns out to be extremely diffuse in character, the density
criterion \cref{eq:rvdw} is not satisfied with any radius.  As a
result, $r_\rho$ is not defined for the employed value of $\epsilon$.

In addition, $r_\rho$ for the $1s^1 3d^1$ configuration of He turns
out to be smaller ($r_\rho = 2.10 a_0$ for spin-unrestricted PBE) than
that for the $1s^1 2p^1$ ($r_\rho = 4.05 a_0$) or $1s^2$ ($r_\rho =
2.68 a_0$) configurations, even though basic physical insight would
say that the $1s^2$ ground state configuration has to be more compact
than those where an electron is excited from the $1s$ shell to the
$2p$ shell, not to mention to the $3d$ shell.

\begin{figure*}
  \includegraphics{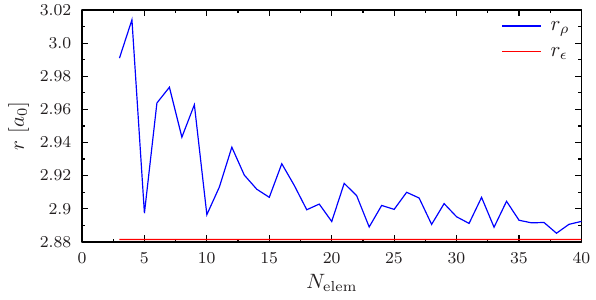}
  \caption{The dependence of the radial estimates $ r_\rho $ and $
    r_\epsilon $ defined by \cref{eq:rvdw,eq:el_count}, respectively,
    on the number of radial elements $ N_\text{elem} $ used in the
    discretization. The calculations are performed for the ground
    state of the hydrogen atom, using unrestricted Hartree--Fock
    theory, and the error in the total energy compared to the
    analytical value $-0.5 E_h$ is smaller than $10^{-7} E_h$ in all
    calculations shown in the figure. The estimates from
    \cref{eq:el_count} agree with the analytical value $r_\epsilon
    \approx 2.8815126965663684390a_0$ to a precision better than one
    part per million. }
  \label{fig:h-radius}
\end{figure*}

We also find that the criterion of \cref{eq:rvdw} is overly sensitive
to the numerical discretization: a finite element calculation for the
exact ground state of the unconfined hydrogen atom reveals a
significant sensitivity to numerical noise, as shown in
\cref{fig:h-radius}. We find it highly likely that similar issues are
also present in the Gaussian-basis calculations of
\citerefs{Rahm2016_CEJ_14625, Rahm2017_CEJ_4017, Smith2023_JCTC_2064}
that relied on the estimate of \cref{eq:rvdw}: small changes to the
Gaussian basis likely result in considerable changes in the estimated
radius, since the calculations of \cref{fig:h-radius} which all
reproduce the exact ground state energy $-1/2 E_h$ to sub-$\mu E_h$
accuracy exhibit considerable differences in the estimate of the vdW
radius.

Note that the ANO-RCC basis set employed in
\citerefs{Rahm2016_CEJ_14625} and \citenum{Rahm2017_CEJ_4017} produces
an analogous Hartree--Fock energy of $-0.49998374 E_h$ (basis set
obtained from the Basis Set Exchange\cite{Pritchard2019_JCIM_4814} and
calculation performed with ERKALE\cite{Lehtola2012_JCC_1572}), and
thus exhibits a much larger truncation error of $16.3 \mu E_h$ for the
unconfined hydrogen atom. \citeit{Smith2023_JCTC_2064} examined
various Gaussian basis sets and variations in the order of $0.01a_0$
can be observed in their data, as well.

To choose the criterion $\epsilon$ in \cref{eq:el_count}, we match the
two estimates of the vdW radius using the well-known analytical
expression for the ground state of the hydrogen atom
($R_{1s}(r)=2e^{-r}$, $f_{1s}=1$). Performing the matching using
20-digit precision in Maple 2024, the electron density is found to be
$10^{-3}$ electrons/bohr$^3$ at $r_\epsilon \approx
2.8815126965663684390a_0 \approx 1.524$ \AA{}, and evaluating
\cref{eq:el_count} yields the threshold $\epsilon \approx
0.073416683704840394115$ electrons for use in determining the radius
from \cref{eq:el_count}.

The data in \cref{fig:h-radius} show that the proposed $r_\epsilon$
estimate of \cref{eq:el_count} is numerically well-behaved, yielding
excellent agreement with the radius determined from the analytical
solution, thus successfully addressing the issues with undefined vdW
radii and numerical stability in the $r_\rho$ estimate of
\cref{eq:rvdw}. This criterion also easily resolves the issue with the
configurations of He, agreeing with physical insight that the $1s^2$
ground state ($r_\epsilon = 2.26 a_0$ for spin-unrestricted PBE) is
more compact than the $1s^1 2p^1$ ($r_\epsilon = 8.07 a_0$) or $1s^1
3d^1$ ($r_\epsilon = 15.48 a_0$) excited states, the latter of which
also show a significant difference in diffuse character.

The $r_\epsilon$ estimates for the states of He are also in excellent
qualitative agreement with the covalent radius estimates $r_{\rm max}$
of \cref{eq:densmax}: both $r_\epsilon$ and $r_{\rm max}$ predict the
size order $1s^2$ < $2p^2$ < $1s^1 2p^1$ < $1s^1 3d^1$, while the
$r_\rho$ estimate of \cref{eq:rvdw} predicts a thoroughly dissimilar
order $1s^1 3d^1$ < $1s^2$ < $1s^1 2p^1$ < $2p^2$.

We note again here that the $1s^1 3d^1$ excited state is so
diffuse in the unconfined He atom that it is likely not converged to
the complete basis set limit in our calculations, as we chose not to
converge the value of the practical infinity $r_\infty$ given that
excited states of the unconfined atom are not the focus of this work.

\subsection{Observed changes in electronic structure} \label{sec:changes}

In the following, we will discuss the key observations from the
systematic analysis of all the atoms H--Xe. We focus the discussion in
the main text on the PBE functional due to its insurmountable
popularity in the literature, but note that the behaviors of various
atoms are qualitatively independent of the functional, as was already
demonstrated for the Ca atom in \cref{sec:ion_radii}. Thorough
analyses for all functionals are available in the SI.

In addition to energies of the configurations as a function of
confinement, we also report the estimated atomic radii for the
configurations of the unconfined atoms with the methodology discussed
in \cref{sec:atomic-radii,sec:radii-results} in the SI.
We note again the limitations of the previously used $r_\rho$ estimate of
\cref{eq:rvdw} that was discussed above in \cref{sec:radii-results}, in
that the reported $r_\rho$ values exhibit numerical noise.
Furthermore, the excitation energies of each configuration relative to
the ground state are also included in the SI for the case of the
unconfined atom.

We note here that we were unable to converge the calculations of some
configurations with the r$^2$SCAN functional. We tentatively attribute this
to the numerical ill-behavedness of the functional\cite{Lehtola2022_JCP_174114}
and to the slow convergence of meta-GGA functionals in general with respect to the
number of radial elements.\cite{Lehtola2023_JCTC_2502}
We have excluded all non-converged calculations from our analysis.

The in-depth analysis of the data included in the SI led us to the
following observations on the behavior of the H--Xe atoms, which can
be grouped into the following classification based on the atom's
position in the periodic table.

\paragraph{H--He}
Similarly to previous studies, we do not observe ground state
crossings for the H or He atoms in the studied range of confinement
radii. The $1s$ orbital remains occupied over the whole domain of
studied confinement radii.  Interestingly, both atoms have
configurations with an occupied $3d$ orbital which are stable with
respect to ionization in the unconfined atom (the configuration has a
lower energy than that of the cation's ground state), even though the
$3d$ orbital is extremely diffuse as $r_\epsilon \approx15a_0$ for the
corresponding configuration.

\paragraph{Li--O}
The valence electrons of the Li--O atoms occupy the $2s$ orbital in
the unconfined atom.  Contrary to the H and He atoms, we observe
ground state crossings where the $2s$ electrons hop to the $2p$
orbital as the confinement radius is decreased; for the Li atom this
happens already at $r_c=3a_0$, but the location of the crossover point
$r_c$ decreases in increasing charge and occurs for the O atom at
$r_c=a_0$.

The electron shift for Li was also predicted by
\citeit{Rahm2019_JACS_10253}, while the $2s^{2} \to 2p^{2}$ transition
of the C atom was reported by \citeit{Pasteka2020_MP_1730989}. As far
as we are aware, the analogous transitions for the Be, B, N, and O
atoms have not been discussed in previous literature.

We further note that the Li and O atoms have configurations
with an occupied $3d$ orbital that are stable with respect to
ionization in the unconfined atom, although the $3d$ orbital is again
extremely diffuse as $r_\epsilon\approx15a_0$,
as in the case of H and He.

\paragraph{F}
The valence electrons in the F atom occupy the $2s$ and $2p$ orbitals.
One of the $2s$ electrons hops to the $2p$ orbital at $r_c=a_0$, which
represents extreme confinement, thus fully filling the $2p$ orbital.
This electron transition does not appear to have been previously
reported in the literature.  Similarly to Li and O, F has a
configuration with an occupied $3d$ orbital already in the unconfined
atom, but the orbital is equally diffuse.

\paragraph{Ne}
Confirming previous studies, the Ne atom does not exhibit ground state crossings in the studied
range of confinement radii.  We note that also Ne has a configuration
with an occupied $3d$ orbital which is stable with respect to
ionization in the unconfined atom.

\paragraph{Na--Mg}
The valence electrons of Na and Mg occupy the $3s$ orbital, which
becomes unfavourable in confinement, like the $2s$ orbital in Li and
Be. While in Li and Be the electrons hopped from $2s$ to $2p$, the
$3s$ electrons of Na and Mg hop to the $3d$ orbital around $r_c
\approx 2 a_0$, instead, even though the $3d$ orbital has a larger
extent than the $3p$ orbital in the unconfined atom, and the $3d$
orbital is not a valence orbital even in the Mg atom without
confinement.\cite{Fernandez2020_CEJ_14194} These electron transitions,
which occur at relatively strong confinement, have not been previously
reported in the literature to our knowledge.

We observe that the [Ne]$4f$ configuration of the Na atom is stable
with respect to ionization in the unconfined atom, and flips below the
initial ground state in extreme confinement (close to $r_c=a_0$).  The
$4f$ orbital is highly diffuse in the unconfined atom ($r_\epsilon \approx 24 a_0$).
Also the state [Ne]$3d^14f^1$ of Mg flips below the initial ground state close to $r_c=a_0$.

\paragraph{Al--Ar}
The valence electrons of the unconfined Al--Ar atoms occupy the $3s$
and $3p$ orbitals.  No electron shifts have been reported in
previous studies for these atoms.  However, we see that the $3p$
orbital becomes unfavourable in confinement, gradually transferring
all electrons to the $3d$ orbital when the confinement radius is
decreased, the transition occurring between $r_c=2a_0$ and
$r_c=1.5a_0$ (strong confinement). As the confinement radius is
further decreased, both the $3s$ electrons follow.  The behavior is
thus different from that observed above for Li--Ne, likely because the
$3d$ orbital is too high to be accessible in the lighter atoms.

\paragraph{K--Ca}
Confirming the results of previous studies,\cite{Rahm2019_JACS_10253,
  Connerade2000_JPBAMOP_251, Guerra2009_AQC_1} analogously to Na and
Mg, the $4s$ valence electrons of the K and Ca atoms hop to the $3d$
orbital around $r_c\approx4.5a_0$.  The $3d$ orbital is thus already
easily accessible, and its relevance in chemistry has been well
established in the literature:\cite{Pyykkoe1979_JCSFT2_1256,
  Wu2018_S_912, Fernandez2020_CEJ_14194, Zhou2021_ACR_3071,
  Liu2023_CS_4872, Cui2024_CEJ_202400714, Liu2024_JACS_16689} Ca is
sometimes referred to in the literature as an ``honorary transition
metal''\cite{Liu2024_JACS_16689} due to its covalent binding like a
transition metal,\cite{Wu2018_S_912, Fernandez2020_CEJ_14194,
  Zhou2021_ACR_3071, Cui2024_CEJ_202400714} following similar work
performed earlier on the heavier analogoues Cs and Ba which are not
considered in this work.\cite{Gagliardi2003_TCA_205}

Further novel ground state crossings happen close to $r_c=a_0$
(extreme confinement), where the [Ar] core configuration is opened up
and all the $3s$ and $3p$ electrons also move to the $3d$ orbital.  We
observe that both atoms have a configuration with an occupied $3d$
orbital that is bound relative to ionization in the unconfined atom.

\paragraph{Sc--Ni}
The $3d$ transition metals feature strong competition between the $4s$ and
$3d$ orbitals, which can be observed from the small excitation
energies in the unconfined atom.  The states [Ar]$3d^n$, $n\leq10$ are
stable with respect to ionization already in the unconfined atom,
which explains why ground state crossings happen already in weak
confinement.

Confirming previous findings,\cite{Rahm2019_JACS_10253,
  Connerade2000_JPBAMOP_251} the $4s$ electrons shift to the $3d$
orbital around $r_c=4a_0$ in the Sc--Ni atoms, similarly to the
observations made above for K and Ca.  However, here we observe that
the Sc and Ti atoms exhibit additional ground state crossings in
extreme confinement (between $r_c=a_0$ and $r_c=2a_0$), where some
$3s$ and $3p$ electrons also shift to the $3d$ orbital.

\paragraph{Cu--Zn}
The unconfined Cu and Zn atoms already have a fully filled $3d$
orbital in their ground states, thus the $4s \rightarrow 3d$ shift
observed earlier in the $3d$ transition metal series cannot happen for
these atoms. The $4s$ electrons hop to the $4f$ orbital in extreme
confinement (around $r_c=1.5a_0$), instead, which does not appear to
have been previously reported in the literature.

\paragraph{Ga--Kr}
The ground state configurations of the unconfined Ga--Kr atoms have
filled $4s$ and $3d$ orbitals and a variably filled $4p$ orbital. The
$4s$ and $4p$ electrons hop to the $4f$ orbital as the confinement
radius is decreased (between $r_c=1.5a_0$ and $r_c=a_0$), leading to a
[Ar]$3d^{10}4f^n$ ground state in extreme confinement. The favouring
of the $4f$ orbital over the $4s$ and $4p$ orbitals in the confined Kr
atom has been previosly reported by \citeit{Garza2005_JCS_379}.

\paragraph{Rb--Sr}
Like the alkali and earth alkali atoms in the previous period, the
valence $5s$ electrons of the Rb and Sr atoms hop to the $4d$
orbital already around $r_c=5a_0$ confirming the previous observations
of \citeit{Guerra2009_AQC_1} and \citeit{Rahm2019_JACS_10253}.
However, unlike the lighter analogues K and Ca, we also see a $4d
\rightarrow 4f$ shift when the confinement radius is further
decreased to $r_c\approx1.2a_0$, which represents extreme
confinement. The $4s$ and $4p$ electrons also shift to the $4f$
orbital below $r_c=1.2a_0$, which, to the best of our knowledge,
has not been reported in the literature so far.

\paragraph{Y--Rh}
The valence electrons of the unconfined Y--Rh atoms occupy the $5s$
and $4d$ orbitals in their ground states. The $5s$ valence orbital
becomes unfavourable in confinement, and the electrons hop to the $4d$
orbital at confinement radii varying from $r_c=6.1a_0$ to
$r_c=4.2a_0$, that is, already in weak confinement, in agreement with
previous literature.\cite{Rahm2019_JACS_10253,
  Connerade2000_JPBAMOP_251} Again, like for Cu--Sr, we observe
further electron shifts to the $4f$ orbital, and the $4d$ orbital
is partially depopulated after the $5s$ orbital in confinement radii
below $r_c=1.5a_0$, which already represents extreme confinement.  Y
and Zr shift electrons from the $4p$ core orbital to the $4f$
orbital close to $r_c=a_0$.  These additional electron shifts do
not appear to have been reported in previous studies.

\paragraph{Pd}
It is well known that the unconfined Pd atom has a particularly stable
ground state ([Kr]$4d^{10}$).\cite{Rahm2019_JACS_10253,
  Connerade2000_JPBAMOP_251} We observe no ground state crossings for
Pd in the studied range of confinement radii.

\paragraph{Ag--Cd}
Ag and Cd already have a filled $4d$ orbital, while we observe
previously unexplored electron shifts where $5s$ electrons hop to
the $4f$ orbital around $r_c=2a_0$.

\paragraph{In--Xe}
The In--Xe atoms have filled $5s$ and $4d$ orbitals and a $5p$ orbital
with variable filling in the unconfined atom. We observe that the $5p$
electrons start to gradually shift to the $4f$ orbital around
$r_c=2a_0$, after which both the $5s$ electrons follow analogously to
Ga--Kr, which does not appear to have been reported in the literature before now.

\section{Summary and conclusion} \label{sec:summary}

To the best of our knowledge, a robust and systematic study of atoms in confinement has not been hitherto published in the literature.
Such a study both needs to use a suitable numerical method and to consider the changing nature of the ground state as a function of the confinement.
We did not find a single publication that satisfies both these criteria while also considering an appreciable portion of the periodic table instead of just some specific subrows or groups.

We therefore took a fully numerical approach based on the finite element method (FEM) to study atoms in hard-wall confinement.
We chose to represent the atoms with spherically averaged, spin-restricted or spin-polarized densities, and carried out calculations at three levels of density functional approximations: the local density approximation employing the Perdew--Wang (PW92) correlation functional, the generalized-gradient approximation (GGA) employing the Perdew--Burke--Ernzerhof (PBE) exchange-correlation functional, and the meta-GGA approximation employing the r$^2$SCAN exchange-correlation functional.
We compared the ionization energies of unconfined atoms predicted by spin-restricted or spin-polarized calculations to experiment, and unsurprisingly observed the latter to be more accurate, while even the spin-restricted calculations succeed in predicting periodic trends.

We demonstrated the significance of considering confinement induced changes in the electronic structure by comparing ionization radii computed for fixed electron configurations to those obtained in calculations in which the electron configuration can relax as a function of the confinement radius.
Depending on the atom, this relaxation can result in either an increase or a decrease of the ionization radius due to competing effects in the neutral atom and its cation, and the largest differences are seen for the pre-$d$ and early $d$ elements.

As a side result of our study, we also examined various atomic size estimates.
We found severe issues in the van der Waals (vdW) radius estimate proposed by \citeit{Bader1967_JCP_3341}: it is not always defined, it has severe issues with numerical stability even with huge numerical basis sets, and its predictions do not agree with basic physical insights for the $1s^2$ ground and $1s^1 2p^1$ and $1s^1 3d^1$ excited configurations of the He atom.
We proposed a related estimate which is always defined, does not suffer from severe numerical problems like the aforementioned estimate, and also predicts the aforementioned states of the He atom to be increasingly diffuse in character, in agreement with the covalent radius estimate of \citeit{Slater1930_PR_57}.
We expect our proposed vdW radius estimate to be useful in future work.

We then carried out a systematic study of the total energies of the ground and low lying excited states of the H--Xe atoms as well as their monocations as a function of the confinement radius $r_c$, paying special focus on how the ground state evolves as a function of confinement.
We have confirmed electron shifts made in previous studies on confined atoms, but also made novel observations.
We summarize our analysis of the calculations as follows.
\begin{itemize}
\item Ground state crossings occur for the majority of the elements.
\item Valence $s$ electrons are highly unfavoured under strong
  confinement: already the Li and Be atoms show a $2s \to 2p$
  electron shift.
\item Shifts from the $ns$ and $np$ to the $3d$ orbital are observed for the elements of the second and third periods.
\item Shifts to the $4f$ orbital can be observed for Cu onwards in strong confinement.
 \end{itemize}

Importantly, we saw that the dissimilar density functional approximations were in excellent agreement in calculations of ionization radii, and that the qualitative behavior of the confined atoms is similar for dissimilar density functionals.
We are therefore confident that our results are not artefacts of the employed level of theory, and that similar results would be obtained even with high-level \textit{ab initio} calculations.

Like most studies on confined atoms, the present study was limited to the use of integer occupation numbers for the $s$, $p$, $d$, and $f$ shells to maximize physical interpretability.
However, as we discussed in \cref{sec:janak}, the use of non-integral occupation numbers could be useful for transition metals, for instance,\cite{Slater1969_PR_672, Kraisler2010_PRA_42516, Lehtola2020_PRA_12516} and we hope to revisit such calculations with state-of-the-art optimization algorithms such as optimal damping\cite{Cances2001_JCP_10616} in future work.

Our study was limited to the H--Xe atoms, because relativistic effects are well known to be essential for heavy elements.\cite{Pyykkoe2012_ARPC_45, Pyykkoe2012_CR_84}
Relativistic calculations of atoms in confinement should be feasible using the same approach, and such calculations might be carried out with the recent implementation of \citeit{Certik2024_CPC_109051}, for example.

\subsection{Connections to basis set design}

It is interesting to draw a parallel here to basis set design, since
the application of confinement potentials in the construction of
numerical atomic orbital (NAO) basis sets as originally proposed by by
\citeit{Averill1973_JCP_6412} was the original motivation of our work.
While \citeit{Averill1973_JCP_6412} employed a finite potential
barrier, \citeit{Sankey1989_PRB_3979} proposed building NAO basis sets
using a hard-wall potential to enforce the locality of the resulting
basis set.  In later works, Sankey and coworkers refer to the arising
orbitals as ``fireballs'' due to their excited nature relative to the
unconfined atom.\cite{Lewis2001_PRB_195103, Lewis2011_PSSB_1989} The
technique of \citeit{Sankey1989_PRB_3979} has been later used by many
other authors as well, such as \citeit{SanchezPortal1997_IJQC_453},
\citeit{Basanta2007_CMS_759}, \citeit{Kenny2009_CPC_2616}, and
\citeit{Nakata2020_JCP_164112}.

The demands for various chemical elements' atomic orbital basis sets
for electronic structure calculations\cite{Davidson1986_CR_681,
  Jensen2013_WIRCMS_273, Hill2013_IJQC_21} are inherently tied to the
elements' chemistry.  Importantly, the relevance of distinct angular
momentum $l$ orbitals for atoms is expected to be a continuous
function: as we move to heavier elements, we first see an increasing
importance of $p$ functions, followed by increasing importance of $d$
functions, $f$ functions and so on. Unlike the occupied orbitals in an
SCF calculation on an atom, the importance of polarization functions
in a basis set is not a step function for which we can certainly say
whether a specific element requires $s$, $p$, $d$, $f$, etc. functions,
or not.

This continuity is a challenge when predicting the relevance of
distinct orbitals to chemical bonding. The advantage of our arguably
simple model is that we are able to highlight these chemical
characteristics of the studied atoms, that is, the \emph{relative}
importance of the electron configurations as a function of
confinement.  The physicality of the model is therefore obvious. For
example, we saw in this work that the simple hard-wall confinement
model predicts the importance of the $3d$ orbital in K and Ca, which
is obvious from the large transition radius $r_c \approx 4.5 a_0$.

In contrast, the $3d$ orbital is so high in the second-row $3p$ block
that electron shifts to it only occur in strong confinement. Yet, it
is by now well known\cite{Curtiss1990_JCP_2537, Curtiss1991_JCP_7221,
  Bauschlicher1995_CPL_533, Martin1998_JCP_2791, Martin1999_CPL_271,
  Dunning2001_JCP_9244} that tight $d$ functions need to be included
in the basis set for the second-row $p$ block elements in polyatomic
electronic structure calculations, as this inclusion typically results
in significant increases of atomization energies: for example, the
atomization energies of sulfur containing molecules such as \ce{SO2}
and \ce{SO3} typically increase by tens of kcal/mol upon the addition
of $3d$ functions. Perchloric acid (\ce{HClO4}) and dichlorine
heptoxide (\ce{Cl2O7}) represent even more extreme cases, with
documented increases of the atomization energy by 50 and 100 kcal/mol,
respectively.\cite{Martin2006_JMST_19}

\citeit{Martin2006_JMST_19} found a physical interpretation for this
effect: the $3d$ Rydberg orbital on the chlorine atom(s) can accept
back-bonding contributions from the oxygen
atoms. \citeit{Mehta2023_JPCA_2104} also point out that the $3d$
orbital comes down in energy when the atom becomes more positively
charged. Indeed, atomic orbitals determined for cations are commonly
used to introduce additional flexibility in NAO basis sets.\cite{Delley1990_JCP_508}
\citeit{Mehta2023_JPCA_2104} also made the discovery that the fourth-row
$5p$ block analogously exhibits a heightened importance of $4f$ Rydberg
orbitals.

Based on the above findings, we expect that heavier elements will
exhibit electron shifts to the $4f$ orbital in weaker and weaker
confinement, since the $4f$ orbital will come down in energy as the
nuclear charge is increased.

\begin{acknowledgement}
  We thank professor Pekka Pyykk\"o for comments on the manuscript, and Monica Lindholm for designing the cover graphics for the manuscript.
  H.\AA{}. thanks the Finnish Society for Sciences and Letters for
  funding. S.L. thanks the Academy of Finland for financial support
  under project numbers 350282 and 353749.
\end{acknowledgement}

\begin{suppinfo}
The SI PDF file contains the following data.
Comparison of the ionization energies with the PW92 and r$^2$SCAN functional.
Values of the ionization radii via Janak's theorem for both the fixed and relaxed configuration approaches.
Detailed analysis of the evolution of the ground state of the H--Xe atoms, and state crossings with the ground state configuration of the unconfined atom.
Estimated atomic sizes and excitation energies relative to the ground state for the studied configurations in the unconfined atom.
Plots of the total energy of the low lying states of the H--Xe atoms and their monocations as a function of the confinement radius.
All of the above sets of results are given both for the spin-restricted and spin-polarized formalisms for all the functionals considered in this work: PW92, PBE, and r$^2$SCAN.
The raw energies for all calculations are included in comma separated values (CSV) format in a compressed archive.
\end{suppinfo}

\begin{tocentry}
\includegraphics{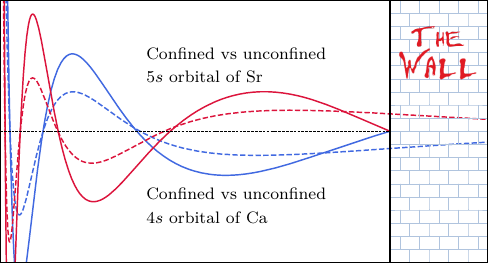}
\end{tocentry}

\bibliography{citations,solid}

\end{document}